\documentclass[12pt]{article}

\usepackage[left=1.5cm,top=2.cm,right=1.5 cm,bottom=2.2cm]{geometry}
\usepackage{graphicx}
\usepackage{amsfonts}
\usepackage[T1]{fontenc}
\usepackage{lmodern}
\usepackage{amsmath}
\usepackage{color}
\usepackage{epsfig}
\usepackage{comment}
\usepackage{amssymb}

\numberwithin{equation}{section}

\def\gtwid{\mathrel{\raise.3ex\hbox{$>$\kern-.75em\lower1ex\hbox{$\sim$}}}}
\def\ltwid{\mathrel{\raise.3ex\hbox{$<$\kern-.75em\lower1ex\hbox{$\sim$}}}}
\def\square{\kern1pt\vbox{\hrule height 1.2pt\hbox{\vrule width 1.2pt\hskip 3pt
   \vbox{\vskip 6pt}\hskip 3pt\vrule width 0.6pt}\hrule height 0.6pt}\kern1pt}

\begin{document}

\begin{titlepage}

\begin{flushright}
Date: \today
\end{flushright}

\vskip .5cm

\begin{center}
{\bf\Large  Gravitational wave signals in an Unruh-DeWitt detector}
\end{center}

\vskip .5cm

\begin{center}
\bf   Tomislav Prokopec$^{\diamondsuit}$
\end{center}

\vskip .5cm

\begin{center}
{
Institute for Theoretical Physics, Spinoza Institute  \& EMME$\Phi$ \\
Utrecht University, Princetonplein 5,
3584 CC Utrecht, The Netherlands \\
}
\end{center}

\begin{center}
{\bf Abstract}
\end{center}

We firstly generalize the massive scalar propagator for planar gravitational waves
propagating on Minkowski space obtained recently in Ref.~\cite{vanHaasteren:2022agf}.
We then use this propagator to study the response of a freely falling Unruh-DeWitt detector 
to a gravitational wave background. 
We find that a freely falling detector completely cancels the effect of the deformation of the invariant 
distance induced by the gravitational waves, such that the only effect comes from an increased average size of scalar field vacuum fluctuations,
the origin of which can be traced back to the change of the surface in which the gravitational waves fluctuate. 
The effect originates from the quantum interference between propagation on off-shell detector's trajectories
which probe different spatial gravitational potential induced by the gravitational backreaction from gravitational waves, 
and it is therefore purely quantum.
When resummed over classical graviton insertions,
gravitational waves generate cuts on the imaginary axis of the complex $\Delta \tau$-plane
(where $\Delta \tau = \tau-\tau'$ denotes the difference of proper times),
and the discontinuity across these cuts is responsible for a continuum of energy transitions
induced in the Unruh-DeWitt detector. Not surprisingly, 
we find that the detector's transition rate is exponentially suppressed with increasing energy 
and the mass of the scalar field. What is surprising, however, is that the transition rate is a
{\it non-analytic function} of the gravitational field strain. 
This means that, no matter how small is the gravitational field amplitude,
expanding in powers of the gravitational field strain cannot approximate well 
the detector's transition rate. 
We present numerical and approximate analytical results for the detector's transition rate 
both for circularly polarized and for polarized monochromatic, unidirectional, gravitational waves.

\begin{flushleft}
\end{flushleft}

\vskip .5cm

\begin{flushleft}
$^{\diamondsuit}$ e-mail: T.Prokopec@uu.nl \\
\end{flushleft}

\end{titlepage}


\section{Introduction}
\label{Introduction}

In this work we calculate the response of a freely falling Unruh-DeWitt detector 
which couples to a massless or massive scalar field fluctuating 
in the presence of planar gravitational waves propagating on Minkowski spacetime. 
This work builds on earlier studies~\cite{Garriga:1990dp,Jones:2016zqw, Jones:2017ejm,Siddhartha:2019yjm,Xu:2020pbj,Chen:2021bcw,vanHaasteren:2022agf},
which address some aspects of the problem of how planar gravitational waves affect scalar fields.
In particular the authors of Ref.~\cite{Chen:2021bcw} investigate the response of freely-falling 
and accelerating Unruh-DeWitt detectors~\cite{Unruh:1976db,Hawking:1979ig,Birrell:1982ix}
in the presence of gravitational waves. 
In this work we generalize their analysis by using the propagator recently obtained in 
Ref.~\cite{vanHaasteren:2022agf}.
For simplicity, we do not analyze here the response of the detector moving along noninertial trajectories.

\bigskip
{\bf The model.}
In this work we consider a real, self-interacting scalar field  $\phi(x)$ whose 
action and Lagrangian are,
\begin{equation}
 S[\phi]=\int d^Dx\sqrt{-g}\,{\cal L}_\phi
 \,,\qquad {\cal L}_\phi = -\frac12 (\partial_\mu\phi)(\partial_\nu\phi) g^{\mu\nu}
 -\frac{m^2}{2}\phi^2
         -\frac{\lambda}{4!}\phi^4
         \,,\quad
\label{scalar field action}
\end{equation}
where $g={\rm det}[g_{\mu\nu}]$, $g^{\mu\nu}$ is the inverse of the metric tensor $g_{\mu\nu}$,
$m$ is the field's mass and $\lambda$ is the self-interaction coupling strength.
We work in natural units in which $c=1$, but keep the dependence on $\hbar$ explicit.
This means that the dimension of the field $\phi$ and the mass $m$ is ${\rm m}^{-1}$,
and $\lambda$ is dimensionless. To restore the physical dimension of $m$, one ought to rescale it as, $m\rightarrow mc/\hbar$.

\subsection{Gravitational waves}
\label{Gravitational waves}

We are interested in understanding the effects of gravitational waves on scalar fields.
A convenient representation for a gravitational wave background is,
 \begin{equation}
g_{\mu\nu}(x)= \eta_{\mu\nu}+ h_{\mu\nu}(x)
\,,
 \label{metric perturbation}
 \end{equation}
where $h_{\mu\nu}(x)$ is a perturbation of the metric tensor
$g_{\mu\nu}$ around flat Minkowski space, characterized by
Minkowski metric $\eta_{\mu\nu}$, which is in Cartesian coordinates 
a $D\times D$ diagonal matrix of the form, 
$\eta_{\mu\nu}={\rm diag}(-1,1,1,\dots)$.
In the traceless-transverse gauge (in which the gravitational field perturbation $h_{\mu\nu}$ 
is gauge invariant to linear order in the gravitational field), 
planar gravitational waves moving in the $x^{D-1}$  direction ($x^{D-1}\rightarrow z$ when $D=4$) 
satisfy $h_{0\mu} = 0$ and 
 \begin{equation}
 h_{ij}(u) = \left(\begin{array}{ccccc}
                        h_{xx}(u) & h_{xy}(u) & \cdots & h_{xD-2}(u) & 0\cr
                         h_{xy}(u) & h_{yy}(u)  & \cdots & h_{yD-2}(u) & 0\cr
                         \vdots & \vdots & \vdots & \cdots & \vdots \cr
                h_{xD-2}(u) & h_{yD-2}(u)  & \cdots & h_{D-2D-2}(u) & 0\cr
                        0 & 0 & \cdots &  0 & 0 \cr
                        \end{array}
               \right)
\,, 
 \label{gravitational wave D}
 \end{equation}
 where $u=t-x^{D-1}$ is a lightcone coordinate.
Note that some of the elements of $h_{ij}$ in Eq.~(\ref{gravitational wave D}) 
may vanish. In $D=4$ dimensions this simplifies to planar gravitational waves
with nonvanishing elements in the $xy-$plane. In practical calculations it is often convenient to 
simplify~(\ref{gravitational wave D}) by assuming that nonvanishing elements of $h_{ij}$ are 
in the upper left $2\times 2$ block.
 
In this work we generalize the monochromatic wave background considered
in Ref.~\cite{vanHaasteren:2022agf} to the case when the gravitational wave strain, $h_{ij}=h_{ij}(u)$, 
is characterized by a general function of $u$ propagating in the $x^{D-1}$ direction.
Motivated by the form of gravitational waves emitted by realistic sources, whose wave form can be 
decomposed into the fundamental mode of frequency $\omega_g$, and the higher overtones
(whose frequencies are $n\omega_g$), we shall consider gravitational waves of the form,
 \begin{equation}
 h_{ij}(u) = \sum_{n=1}^\infty h_{ij}^{(n)}\cos(n\omega_g u+\psi_{ij}^{(n)})
\,, 
 \label{gravitational wave D 2}
 \end{equation}
where $h_{ij}^{(n)}$ and $\psi_{ij}^{(n)}$ are the (time independent) gravitational field amplitudes 
and phases of the $n$-th harmonic. Such elliptically polarized gravitational waves are formed by binary 
systems whose components harbor angular momentum and/or strong magnetic 
fields~\cite{Bourgoin:2022ilm,Amaro-Seoane:2022rxf}. It is important to keep in mind that,
even when gravitational waves are emitted as circularly polarized, the perceived amplitudes 
of the $+$ and $\times$ polarizations will differ, unless the source is {\it face on}, {\it i.e.} 
the inclination angle is zero. This means that the only fixed characteristic of observed gravitational waves
is the relative phase difference, $\Delta \psi=\psi_+-\psi_\times =\pm\pi/2$.

The gravitational waves considered here have a phase velocity, $\vec v_{\rm ph} = \hat z$,
 and are often referred to as the positive frequency solutions. 
In addition there are negative frequency gravitational waves, 
with an opposite phase velocity ($\vec v_{\rm ph} = -\hat z$), for which 
$ h_{ij}= h_{ij}(v)$, with $v=t+z$ ($v=t+x^{D-1}$ in the $D$ dimensional case).
Sufficiently close to the gravitational wave source the gravitational wave propagates radially, such that  
in a relatively small spatial volume one can approximate the wave 
by $h_{ij}(u)$.


\section{Scalar propagator}
\label{Scalar propagator} 
 
Variation of the action~(\ref{scalar field action}) gives a Klein-Gordon equation 
satisfied by the scalar field operator $\hat\phi(x)$,
\begin{equation}
\big(\Box - m^2\big)\hat \phi(x)  = 0
\,,
\label{Klein-Gordon}
\end{equation}
where $\Box = \frac{1}{\sqrt{-g}}\partial_\mu\sqrt{-g}g^{\mu\nu} \partial_\nu$ is the d'Alembertian
operator as it acts on a scalar field, and we have neglected in Eq.~(\ref{Klein-Gordon}) the quartic 
self-coupling. 
The positive and negative frequency Wightman functions are defined as 
the following two-point functions,
\begin{eqnarray}
i\Delta^{(+)}(x;x') \!\!&=&\!\! \left\langle\Omega\right|\hat\phi(x)\hat\phi(x')\left|\Omega\right\rangle 
\,,
 \label{positive frequency Wightman function}\\
 i\Delta^{(-)}(x;x') \!\!&=&\!\! \left\langle\Omega\right|\hat\phi(x')\hat\phi(x)\left|\Omega\right\rangle 
\,,
\label{negative frequency Wightman function}
\end{eqnarray}
where $\left|\Omega\right\rangle$ denotes a state of the scalar field, which for simplicity we 
choose to be the vacuum state. When the field operator in Eq.~(\ref{Klein-Gordon}) 
is expanded in terms of the momentum space mode functions, one can reduce the problem
of obtaining the Wightman functions to performing 
the momentum integrals in 
Eqs.~(\ref{positive frequency Wightman function}--\ref{negative frequency Wightman function})
over products of the mode functions. These integrals  
can be performed by a straightforward generalization of the method 
used in Ref.~\cite{vanHaasteren:2022agf}, whose the main steps 
outline in Appendix~A. From 
Eqs.~(\ref{propagator integral D: 5}--\ref{propagator integral D: 7}) it immediately follows, 
\begin{equation}
i\Delta^{(\pm)}(x;x') = \frac{\hbar m^{D-2}}
     {(2\pi)^\frac{D}{2}[\gamma(u)\gamma(u')]^\frac14\sqrt{{\rm det}[{\cal G}^{ij}](u;u')}}
   \frac{K_{\frac{D-2}{2}}\big(m\sqrt{\Delta {\bar x}_{(\pm)}^2}\,\big)}
   {\big(m\sqrt{\Delta {\bar x}_{(\pm)}^2}\,\big)^\frac{D-2}{2}}
\,,
\label{Wightman functions: general Lor viol solution}
\end{equation}
where $K_{\nu}(z)$ denotes the modified Bessel function of the second kind, and
$\Delta {\bar x}_{(\pm)}^2(x;x')$ are the deformed distance functions, which in lightcone coordinates can be written as,
\begin{eqnarray}
\Delta {\bar x}_{(\pm)}^2(x;x') \!\!&=&\!\!
 -(\Delta u\!\mp\!i\epsilon)(\Delta v\!\mp\!i\epsilon)
  +\left(\!\begin{array}{cccc}\Delta x & \Delta y & \cdots  & \Delta x^{D-2} \end{array}\!\right)
     \!\cdot\! \mathbf{\cal G}(u;u')
    \!\cdot\!\left(\!\begin{array}{c}
                         \Delta x\cr 
                        \Delta y\cr
                        \vdots\cr
                        \Delta x^{D-2}\cr
                       \end{array}
                \!\right)
\,,\qquad
\label{Lorentz breaking distance functions: LC coord}
\end{eqnarray}
and in Cartesian coordinates,
\begin{eqnarray}
\Delta {\bar x}_{(\pm)}^2(x;x')   \!\!&=&\!\! -(\Delta t \mp i\epsilon)^2
  +\left(\!\begin{array}{cccc}\Delta x & \Delta y & \cdots  & \Delta x^{D-2} \end{array}\!\right)\!\cdot\! \mathbf{\cal G}(u;u')
    \!\cdot\!\left(\!\begin{array}{c}
                         \Delta x\cr 
                        \Delta y\cr
                        \vdots\cr
                        \Delta x^{D-2}\cr
                       \end{array}
                \!\right)
        + \Delta x_{D-1}^2
\,,\qquad
\label{Lorentz breaking distance functions: Cart coord}
\end{eqnarray}
where $\Delta x^\mu = x^\mu - {x'}^\mu$, where the deformation matrix ${\cal G}_{ij}(u;u')$
is given by,
\begin{eqnarray}
{\cal G}_{ij}(u;u') \!\!&=&\!\! 
               \left(\!\!\begin{array}{cccc}
                        {\cal G}_{xx}(u;u') & {\cal G}_{xy}(u;u') & \cdots &{\cal G}_{x D-2}(u;u') \cr 
                       {\cal G}_{xy}(u;u') &  {\cal G}_{yy}(u;u') & \cdots &{\cal G}_{y D-2}(u;u')\cr
                      \vdots &    \vdots &  \vdots &   \vdots   \cr
                {\cal G}_{x D-2}(u;u') &  {\cal G}_{y D-2}(u;u') & \cdots &{\cal G}_{y D-2}(u;u')\cr
                       \end{array}
                \!\!\right)
\,.\qquad
\label{deformation matrix: position space}
\end{eqnarray}
Note that ${\cal G}_{ij}(u;u')$
is the inverse of the corresponding momentum space deformation matrix 
${\cal G}^{ij}(u;u')$, ${\cal G}^{ik}(u;u'){\cal G}_{kj}(u;u')=\delta^i_{\;j}$. 
The general form of ${\cal G}^{ij}(u;u')$ is,
\begin{eqnarray}
{\cal G}^{ij}(u;u')(u;u') \!\!&=&\!\! \frac{1}{\Delta u}\int_{u'}^u 
               \left(\!\!\begin{array}{cccc} 
                        g^{xx}(\bar u) & g^{xy}(\bar u) & \cdots &  g^{x D-2}(\bar u) \cr 
                        g^{xy}(\bar u) & g^{yy}(\bar u) & \cdots &  g^{y D-2}(\bar u) \cr 
                      \vdots &    \vdots &  \vdots &   \vdots   \cr
                        g^{x D-2}(\bar u) & g^{y D-2}(\bar u) & \cdots &  g^{D-2 D-2}(\bar u) \cr
                       \end{array}
                \!\!\right) {\rm d}\bar u
\,,\qquad
\label{deformation matrix: momentum space}
\end{eqnarray}
where $ g^{ij}(u)$ denote the inverse of $g_{ij}(u)$. For example, for gravitational waves 
oscillating in the $xy$ plane, only the distances in this plane (which we denote by $\perp$)
get deformed, such that the nontrivial elements of the deformation matrix are,
\begin{eqnarray}
{\cal G}^{ij}_\perp(u;u') \!\!&=&\!\! \frac{1}{\Delta u}\int_{u'}^u
               \left(\!\!\begin{array}{cc} 
                        g^{xx}(\bar u) & g^{xy}(\bar u)\cr 
                        g^{xy}(\bar u) & g^{yy}(\bar u) \cr 
                       \end{array}
                \!\!\right) {\rm d}\bar u
\nonumber\\
\!\!&=&\!\! \frac{1}{\Delta u}\int_{u'}^u \frac{1}{\gamma(u)}
               \left(\!\!\begin{array}{cc} 
                        g_{yy}(\bar u) & -g_{xy}(\bar u)\cr 
                        -g_{xy}(\bar u) & g_{xx}(\bar u) \cr 
                       \end{array}
                \!\!\right){\rm d}\bar u
\,,\qquad
\label{deformation matrix: momentum space 2}
\end{eqnarray}
where $\gamma(u)={\rm det}[g_{ij}(u)]$. For monochromatic, circularly polarized 
gravitational waves in linear representation one obtains
(see section~3 of Ref.~\cite{vanHaasteren:2022agf}), 
\begin{eqnarray}
{\cal G}_{ij}^\perp(u;u')
\!\!&=&\!\! \frac{1}{\gamma{\rm det}[{\cal G}_\perp^{ij}](u;u')}\left(\!\!\begin{array}{cc}
                 1 \!+\! h \frac{\sin(\omega_g u)-\sin(\omega_g u')}{\omega_g\Delta u} &
                        -h\frac{\cos(\omega_g u)-\cos(\omega_g u')}{\omega_g\Delta u} \cr
                         -h\frac{\cos(\omega_g u)-\cos(\omega_g u')}{\omega_g\Delta u} &  
                  1 \!-\! h\frac{\sin(\omega_g u)-\sin(\omega_g u')}{\omega_g\Delta u}\cr
                  \end{array}\!\!\right)
\,,
\label{inverse rotation matrix}\\
  {\rm det} \left[{\cal G}_\perp^{ij}(u;u')\right]
 \!\!&=&\!\!  \frac{1}{\gamma^2}\bigg[1\!-\!h^2j^2_0\Big(\frac{\omega_g\Delta u}{2}\Big)\bigg]
\,,\quad
\label{determinant upsilon: linear rep nonpol}
\end{eqnarray}
where $j_0(z)=\sin(z)/z$ is the spherical Bessel function,
$\gamma(u)={\rm det}[g_{ij}(u)]=1-h^2$ is time independent, 
$h=h_+=h_\times$, and $h_+=h_{xx}=-h_{yy}$ and $h_\times = h_{xy}$ are 
the amplitudes of the two polarizations.
The matrix $\mathbf{\cal G}_{ij}^\perp(u;u')$ deforms distances
in position space according to 
Eqs.~(\ref{Lorentz breaking distance functions: LC coord}--\ref{Lorentz breaking distance functions: Cart coord}).

On the other hand, for singly polarized monochromatic waves fluctuating in the $xy$-plane one obtains
for the $(+)$-polarized waves ($h_+\neq 0$, $h_\times=0$),
\begin{equation}
{\cal G}^{ij}_\perp = \frac{2}{\omega_g\Delta u}\frac{1}{\sqrt{1\!-\!h_+^2}}
\left\{\left(\!\!\begin{array}{cc}          
\text{arctan}\bigg[\sqrt{\frac{1-h_+}{1+h_+} }\tan\Big( \frac{\omega_g u}{2} \Big)
                   \bigg] \;\;& 0\cr
                      0  
\;\;& \text{arctan}\bigg[\sqrt{\frac{1+h_+}{1-h_+} }
                        \tan\Big( \frac{\omega_g u}{2}\Big)\bigg] \cr
                                      \end{array}\!\!\right)
\!-\! (u\rightarrow u')\right\}
,
\label{Upsilon matrix: plus lin}
\end{equation}
and for the $(\times)$-polarized waves ($h_+= 0$, $h_\times\neq0$),
\begin{equation}
{\cal G}^{ij}_\perp   = \frac{1}{\omega_g\Delta u}\frac{1}{\sqrt{1\!-\!h_\times^2}}
\left\{\left(\!\!\begin{array}{cc}          
\text{arctan}\bigg[\frac{1}{\sqrt{1\!-\!h_\times^2} }\tan(\omega_g u)\bigg]
                \;&   - \text{arctan}\bigg[\frac{h_\times}{\sqrt{1-h_\times^2} } \sin(\omega_g u)
                   \bigg] \cr
                    -\text{arctan}\bigg[\frac{h_\times}{\sqrt{1-h_\times^2} }\sin(\omega_g u)
                   \bigg]  
\;& \text{arctan}\bigg[\frac{1}{\sqrt{1-h_\times^2} }\tan(\omega_g u)\bigg] \cr
                                      \end{array}\!\!\right)
                                     \! -\! (u\rightarrow u')\right\}
,
\label{Upsilon matrix: plus lin}
\end{equation}
respectively.

\bigskip 

From the Wightman functions~(\ref{Wightman functions: general Lor viol solution})
one can easily construct the Feynman propagator. In lightcone coordinates
we have, 
\begin{eqnarray}
i\Delta_{F,LC}(x;x') \!\!\!&\equiv&\!\!\! \Theta(\Delta u)i\Delta^{(+)}(x;x')
                                      \!+\! \Theta(-\Delta u)i\Delta^{(-)}(x;x')
\nonumber\\
\!\!&=&\!\! 
\frac{\hbar m^{D-2}}{(2\pi)^\frac{D}{2}[\gamma(u)\gamma(u')]^\frac14
 \sqrt{{\rm det}[{\cal G}^{ij}](u;u')}}
   \frac{K_{\frac{D-2}{2}}\big(m\sqrt{\Delta {\bar x}_{F,LC}^2}\,\big)}
   {\big(m\sqrt{\Delta {\bar x}_{F,LC}^2}\big)^\frac{D-2}{2}}
,\qquad\;\;
\label{Feynman propagator: lightcone coordinates}
\end{eqnarray}
and in Cartesian coordinates,
\begin{eqnarray}
i\Delta_F(x;x') \!\!&\equiv&\!\! \Theta(\Delta t)i\Delta^{(+)}(x;x')
                                      + \Theta(-\Delta t)i\Delta^{(-)}(x;x')
\nonumber\\
\!\!&=&\!\! 
\frac{\hbar m^{D-2}}{(2\pi)^\frac{D}{2}[\gamma(u)\gamma(u')]^\frac14
\sqrt{{\rm det}[{\cal G}^{ij}](u;u')}}
   \frac{K_{\frac{D-2}{2}}\big(m\sqrt{\Delta {\bar x}_{F}^2}\,\big)}
   {\big(m\sqrt{\Delta {\bar x}_{F}^2}\big)^\frac{D-2}{2}}
\,.\qquad
 \label{Feynman propagator: Cartesian coordinates}
 \end{eqnarray}
Both propagators~(\ref{Feynman propagator: lightcone coordinates}--\ref{Feynman propagator: Cartesian coordinates})
are suitable for perturbative studies, the former for the initial
value problem defined on an $u={\rm constant}$ hypersurface,
the latter on a $t={\rm constant}$ hypersurface. However, 
the two $i\epsilon$ prescriptions differ,
\begin{eqnarray}
\Delta {\bar x}_{F,LC}^2(x;x')  
      \!\!&=&\!\! -(|\Delta u| - i\epsilon) (\pm\Delta v-i\epsilon)\!+\!\|\Delta\vec{\bar x}_\perp\|^2
\,,
\nonumber\\
\Delta {\bar x}_{F}^2(x;x')  
\!\!&=&\!\! -(|\Delta t| - i\epsilon)^2\!+\!\|\Delta\vec{\bar x}\|^2
\,.\qquad\quad
 \label{Feynman propagator: i epsilon prescriptions}
 \end{eqnarray}
That means that the imaginary parts of the propagators differ.
Both prescriptions are legitimate, as they are designated to study inequivalent 
perturbative evolution problems.

From 
Eqs.~(\ref{Feynman propagator: lightcone coordinates}--\ref{Feynman propagator: Cartesian coordinates}) one easily obtains the corresponding Dyson propagators,
\begin{equation}
i\Delta_{D,LC}(x;x') = \big[ i\Delta_{F,LC}(x;x')\big]^*
\,,\qquad
i\Delta_{D}(x;x') = \big[ i\Delta_F(x;x')\big]^*
\,,\qquad
\label{Dyson propagator}
\end{equation}
which are important for studying time evolution of Hermitian operators in interacting quantum field theories.

{\bf One-loop results.} In what follows we briefly summarize the one-loop calculations
from Ref.~\cite{vanHaasteren:2022agf}
for the generalized gravitational waves of the form~(\ref{gravitational wave D}). 

For the one-loop effective action calculation and one-loop scalar mass 
induced by the scalar self-interaction, one needs 
the coincident propagator~(\ref{Feynman propagator: lightcone coordinates}--\ref{Feynman propagator: Cartesian coordinates}), which is of the same form as in Eq.~(4.4) of Ref.~\cite{vanHaasteren:2022agf},
\begin{equation}
 i\Delta_F(x;x) 
 =   \frac{\hbar m^{D-2}}{(4\pi)^{D/2}\sqrt{\gamma(u){\rm det}[{\cal G}^{ij}(u;u)]}}
                                  \Gamma\Big(1-\frac{D}{2}\Big)
\,,
\label{coincident propagator}
\end{equation}
where $\gamma(u) = {\rm det}[g_{ij}]$ and ${\rm det}[{\cal G}^{ij}(u;u)]$
 is the determinant of 
the ${\cal G}^{ij}(u;u)$ matrix in Eq.~(\ref{deformation matrix: momentum space}) 
evaluated at spacetime coincidence.
Applying the l'Hospital rule to Eq.~(\ref{deformation matrix: momentum space 2}) yields,
\begin{eqnarray}
{\cal G}^{ij}(u;u) \!\!&=&\!\!
               \left(\!\!\begin{array}{cccc} 
                        g^{xx}(u) & g^{xy}(u) & \cdots &  g^{x D-2}(u) \cr 
                        g^{xy}(u) & g^{yy}(\bar u) & \cdots &  g^{y D-2}(u) \cr 
                      \vdots &    \vdots &  \vdots &   \vdots   \cr
                        g^{x D-2}(u) & g^{x D-2}(u) & \cdots &  g^{D-2 D-2}(u) \cr
                       \end{array}
                \!\!\right) \equiv g^{ij}(u)
\,,\qquad
\label{deformation matrix: coincidence}
\end{eqnarray}
from which it immediately follows that, 
${\rm det}[{\cal G}^{ij}](u;u)={\rm det}[g^{ij}(u)]=1/\gamma(u)$,
and therefore, 
\begin{equation}
 \gamma(u){\rm det}[{\cal G}^{ij}(u;u)] = 1
 \,.
\label{product gamma Upsilon}
\end{equation}
This shows that both, the one-loop effective action and the one-loop scalar mass 
reduce to those of Minkowski space in Eqs.~(4.12) and~(4.15) of 
Ref.~\cite{vanHaasteren:2022agf}.

The calculation of the one-loop energy momentum tensor is more involved,
but the procedure is the same as for the polarized gravitational waves in linear representation
in section~5 of Ref.~\cite{vanHaasteren:2022agf}, and the resulting 
renormalized energy momentum tensor is identical in form as in Eqs.~(5.29-5.30)
of Ref.~\cite{vanHaasteren:2022agf},
\begin{eqnarray}
\langle\Omega | T^*[ \hat T^{\rm ren}_{\mu\nu}(x)]|\Omega\rangle
      \!\!&=&\!\! - \frac{\hbar m^4}{64\pi^2}
                                 \left[\ln\left(\frac{m^2}{4\pi \mu^2}\right)\!+\!\gamma_E\!-\!\frac{3}{2}
                                   \right]g_{\mu\nu}(u)
 - \frac{\hbar m^2}{96\pi^2}\left[
            \log\left(\frac{m^2}{4\pi\mu^2}\right)
            \!+\!\gamma_E\!-\!1\right] 
            G_{\mu\nu}(u)
\,,\qquad\;\;
\label{ren energy momentum tensor}
\end{eqnarray}
where $G_{\mu\nu}(u)$ is the classical Einstein tensor associated with the metric
in Eqs.~(\ref{metric perturbation}--\ref{gravitational wave D}).
The counterterms needed to renormalize~(\ref{ren energy momentum tensor}) 
are generated by the cosmological constant action and the Hilbert-Einstein action,
as detailed in Ref.~\cite{vanHaasteren:2022agf}.
Upon recalling that gravitational waves carry a classical 
(Lifshitz) energy-momentum tensor, $T_{\mu\nu}^{\rm class} = -G_{\mu\nu}(u)/(8\pi G)$,
the result in Eq.~(\ref{ren energy momentum tensor}) can be intuitively understood
as the one-loop  scalar matter energy momentum tensor induced 
by the leading quantum response of the massive scalar field to passing gravitational waves.
The result~(\ref{ren energy momentum tensor}) cannot be directly compared with 
that of Ref.~\cite{Gibbons:1975jb},
where the author considered the one-loop energy-momentum tensor of a massive scalar field
observed in a distant future (in which the vacuum state reduces to that of Minkowski space)
and argued that the one-loop energy momentum tensor is identical
to that in Minkowski vacuum. Note also that, if the gravitational wave amplitude is adiabatically 
switched off, the result in Eq.~(\ref{ren energy momentum tensor}) reduces to 
the trivial (Minkowski vacuum) result of Ref.~\cite{Gibbons:1975jb}.


\section{Unruh-DeWitt detector}
\label{Unruh-DeWitt detector} 

In this section we study the response of a  freely falling Unruh-DeWitt 
detector~\cite{Unruh:1976db,Hawking:1979ig,Birrell:1982ix}
moving in the background of gravitational waves which propagate in the $z$-direction. This work
generalizes the analysis of Ref.~\cite{Chen:2021bcw}.

\bigskip
{\bf Freely falling observers.}
The line element for the problem at hand can be written from Eq.~(\ref{metric perturbation}) as, 
\begin{equation}
{\rm d}s^2 = -{\rm d}t^2 + {\rm d}x^i g_{ij}(u){\rm d}x^j
              = -{\rm d}u{\rm d}v + {\rm d}x^i g^\perp_{ij}(u){\rm d}x^j
\,,
\label{line element}
\end{equation}
where $g^\perp_{ij}(u)$ is a $(D-2)\times (D-2)$ dimensional symmetric metric tensor
(in $D=4$ it reduces to a $2\times 2$ dimensional symmetric metric).
Useful killing vectors are, ${K}_v=\partial_v$ and $K_i = \partial_i\;(i=1,2,\cdots , D-2)$, 
from which one obtains the corresponding conserved 
momenta,~\footnote{Recall that each Killing vector $K$ 
obeys a Killing equation, $\nabla_{(\mu}K_{\nu)}=0$, and 
generates a conserved quantity, $P= K_\mu \frac{{\rm d}x^\mu}{{\rm d}\lambda}$.}
\begin{equation}
P_v = -(K_v)_\mu\frac{{\rm d}x^\mu}{{\rm d}\tau} = \frac12\frac{{\rm d}u}{{\rm d}\tau}
\,,\qquad 
P_i  
    = (K_i)_j\frac{{\rm d}x^j}{{\rm d}\tau} =g^\perp_{ij}(u)\frac{{\rm d}x^j}{{\rm d}\tau}
\,,\qquad (i,j=1,2,\cdots , D-2)
\,,
\label{conserved momenta}
\end{equation}
where (for a later convenience) we chose the geodesic time $\lambda=\tau$ 
to be the proper time $\tau$, defined by ${\rm d}\tau^2=-{\rm d}s^2$.
Upon inserting these equations into the line element~(\ref{line element}) and dividing by $-{\rm d}\tau^2$ one obtains, 
\begin{equation}
1 = \frac{{\rm d}u}{{\rm d}\tau}\frac{{\rm d}v}{{\rm d}\tau}
               - \frac{{\rm d}x^i}{{\rm d}\tau} g^\perp_{ij}(u)\frac{{\rm d}x^j}{{\rm d}\tau}
    = 2P_v\frac{{\rm d}v}{{\rm d}\tau}  - P_ig_\perp^{ij}(u) P_j
\,.
\label{line element 2}
\end{equation}
This generates the geodesic equation for ${\rm d}v/{\rm d}\tau$,
\begin{equation}
\frac{{\rm d}v}{{\rm d}\tau} =\frac{1}{2P_v}\left( 1 + P_ig_\perp^{ij}(u) P_j\right) 
\,,
\label{geodesic equation for v(u)}
\end{equation}
whose formal solution is, 
\begin{equation}
v(\tau) =v_0+\frac{1}{2P_v}\left(\tau+\frac{P_i}{2P_v}\int_{u_0}^u {\rm d}\bar u g_\perp^{ij}(\bar u) P_j\right) 
\,,\qquad (v_0 \equiv v(0),\;u_0 \equiv u(0))
\,,
\label{geodesic equation for v(u) 2}
\end{equation}
where we made use of, ${\rm d}u/{\rm d}\tau=2P_v$.
Next, one can solve equations~(\ref{conserved momenta}) to obtain, 
\begin{eqnarray}
 u(\tau)  \!\!&=&\!\! u_0+ 2P_v \tau
\,,\qquad  (u_0 \equiv u(0))
\label{conserved momenta 2A}\\
 x^i(\tau)
     \!\!&=&\!\! x^i_0 + \frac{1}{2P_v}\int_{u_0}^u{\rm d}\bar u  g_\perp^{ij}(\bar u) P_j  
     \,,\qquad  (x^i_0 \equiv x^i(0))
\,,
\label{conserved momenta 2B}
\end{eqnarray}
such that Eq.~(\ref{geodesic equation for v(u) 2}) can be also written as, 
\begin{equation}
v(\tau) =v_0+\frac{1}{2P_v}\left(\tau+ P_i (x^i(u)-x^i_0)\right)
\,.
\label{geodesic equation for v(u) 3}
\end{equation}
From Eq.~(\ref{conserved momenta 2A}) we see that one can always replace $\tau$ with $u$,
\begin{equation}
\tau(u) = \frac{u - u_0}{2P_v} 
\,.\qquad 
\label{u as function of tau}
\end{equation}

\bigskip
{\bf Unruh-DeWitt detector.} An Unruh-DeWitt 
detector~\cite{Unruh:1976db,Hawking:1979ig,Birrell:1982ix}
is a detector with a monopolar coupling to a scalar field $\phi$, which can be represented by
the interaction Lagrangian, 
\begin{equation}
{\cal L}_{\rm int} = - g m(t)\phi(x)
\,,\qquad
\label{Lagrangian: UdW detector}
\end{equation}
where $m(t)$ is the monopole moment of the detector and $g$ a coupling constant. 
At the first order of perturbation theory, the transition 
amplitude from the ground state, $ |E_0\rangle \otimes|\Omega\rangle$
(where $|E_0\rangle$ denotes the ground state of the detector with energy $E_0$ 
and $|\Omega\rangle$ denotes the ground state of the scalar field) 
to a state $|E\rangle \otimes|\Psi\rangle$ (where $|E\rangle$ is an excited state of the detector with
energy $E>E_0$), is given by, 
\begin{equation}
{\cal A}(E_0\rightarrow E) = i g\int_{-\infty}^{\infty} \langle E|\hat m(\tau)|E_0\rangle
         \langle \Psi|\hat\phi(x(\tau)){\rm d}\tau|\Omega\rangle
\,,\qquad
\label{Transition rate: UdW detector}
\end{equation}
where $\tau$ is the geodesic time and $x^\mu(\tau)$ parametrizes a geodesic. 
The probability $P$ that the detector transits from $E_0$ to $E$ is obtained by squaring 
the transition amplitude, and summing over all intermediate (excited) states of the field $|\Psi\rangle$, 
resulting in,
\begin{equation}
P = g^2\sum_E|\langle E|\hat m(0)|E_0\rangle |^2{\cal F}(\Delta E) 
\,,\qquad
\label{probability: UdW detector}
\end{equation}
where we made use of, $\hat m(0) = {\rm e}^{-i\hat H_0 \tau}\hat m(\tau) {\rm e}^{i\hat H_0 \tau}$,
denoting the monopole moment evolved back to the initial time $\tau=0$, and 
${\cal F}(\Delta E)$ denotes the response function of the detector given by,
\begin{equation}
{\cal F}(\Delta E) = \int_{-\infty}^{\infty}{\rm d}\tau  \int_{-\infty}^{\infty}{\rm d}\tau'
 {\rm e}^{-i\Delta E(\tau-\tau')} i\Delta^{(+)}\big(x^\mu(\tau);x^\nu(\tau')\big)
\,,\qquad \Delta E= E-E_0
\,.\qquad 
\label{response function: UdW detector}
\end{equation}
Here $ i\Delta^{(+)}\big(x^\mu(\tau);x^\nu(\tau')\big)$ is the positive frequency 
Wightman function~(\ref{Wightman functions: general Lor viol solution}) 
evaluated along the geodesics of the detector,
\begin{equation}
x^\mu = x^\mu(\tau)\,,\qquad x^{\prime\nu} = x^\nu(\tau')
\,.\qquad 
\label{geodesics for UdW detector}
\end{equation}
It is useful to transform the integrals in Eq.~(\ref{response function: UdW detector}) to the relative 
and average proper times, $\Delta\tau = \tau-\tau'$ and $T = (\tau+\tau')/2$,
\begin{equation}
{\cal F}(\Delta E) = \int_{-\infty}^{\infty}{\rm d}T  \int_{-\infty}^{\infty}{\rm d}\Delta \tau
 {\rm e}^{-i\Delta E\Delta \tau} i\Delta^{(+)}\big(x^\mu(T+\Delta\tau/2);x^\nu(T-\Delta\tau/2)\big)
\,,\qquad 
\label{response function: UdW detector 2}
\end{equation}
such that one can define the {\it transition rate} ${\cal R}$ as the rate of detector's transitions,
$E_0\rightarrow E$, per unit time, 
\begin{eqnarray}
{\cal R}(T,\Delta E) \!\!&=&\!\! {\rm lim}_{\Delta T \rightarrow 0}\left[\frac{\Delta {\cal F}(\Delta E) }{\Delta T}\right]
\nonumber\\
   \!\!&=&\!\! \int_{-\infty}^{\infty}{\rm d}\Delta \tau
 {\rm e}^{-i\Delta E\Delta \tau} i\Delta^{(+)}\big(x^\mu(T+\Delta\tau/2);x^\nu(T-\Delta\tau/2)\big)
\,.\qquad 
\label{transition rate: UdW detector 0A}
\end{eqnarray}
 Strictly speaking this transition rate 
is valid only for eternal gravitational waves. Realistic gravitational waves 
are transients with a finite duration $\Delta T$, suggesting that the integration limits for $\Delta \tau$ should be 
placed roughly at $\pm\Delta T/2$. However, due to the oscillatory character of the integrand
(generated by the factor ${\rm e}^{-i\Delta E\Delta \tau}$), which is responsible for
a destructive interference 
at large $\Delta \tau$'s, as long as $\Delta E \Delta T\gg 1$, the finite limits of integration will not significantly affect 
the integral in~(\ref{transition rate: UdW detector 0A}), and thus we shall neglect it in what follows;
for a more comprehensive discussion of this point see Ref.~\cite{Louko:2007mu}.

Now, from Eqs.~(\ref{conserved momenta 2A}--\ref{conserved momenta 2B})
and~(\ref{geodesic equation for v(u) 3}) one easily obtains, 
\begin{eqnarray}
 \Delta u(\tau)  \!\!&=&\!\!  2P_v\Delta\tau
\,,\qquad\qquad \qquad\qquad\quad\;\; \, (\Delta u_0=0)
\,,
\label{geodesic equation Delta u}\\
 \Delta x^i(\tau)
     \!\!&=&\!\! \frac{\Delta u}{2P_v}{\cal G}^{ij}(u;u') P_j  
     \,,\qquad\qquad \quad\,   (i,j=1,2,\cdots,D\!-\!2,\;\Delta x^i_0=0)
\,,
\label{geodesic equation Delta xi}\\
\Delta v \!\!&=&\!\! \frac{\Delta u}{4P_v^2}
\Big[1+P_i{\cal G}^{ij}(u;u') P_j \Big]
\,,\qquad\! (\Delta v_0 =0)
\,,
\label{geodesic equation for Delta v}
\end{eqnarray}
where $\Delta x^\mu_0=0$ follows from the fact that we are considering worldlines of a single particle.
From Eq.~(\ref{geodesic equation Delta u}) we see 
that the conserved momentum $2P_v$ converts a proper time interval $\Delta \tau$ into the coordinate time interval $\Delta u$.
When these are inserted into Eqs.~(\ref{Lorentz breaking distance functions: LC coord}) one obtains,
\begin{eqnarray}
\Delta {\bar x}_{(\pm)}^2(x;x') \!\!&=&\!\!
  -(\Delta u\!\mp\!i\epsilon)\left[
   \frac{\Delta u}{4P_v^2}\big(1+P_i{\cal G}^{ij}(u;u') P_j\big)\!\mp\!i\epsilon\right]
 \nonumber\\
 \!\!&+&\!\!\left(\frac{\Delta u}{2P_v}{\cal G}^{ik}(u;u') P_k \right)
        {\cal G}_{ij}(u;u')\left( \frac{\Delta u}{2P_v}{\cal G}^{jl}(u;u') P_l \right)
 \nonumber\\
      \!\!&=&\!\!        -\frac{(\Delta u\!\mp\!i\epsilon)^2}{4P_v^2} 
                          = -(\Delta \tau\!\mp\!i\epsilon)^2
                          =-\frac{(\Delta t\!\mp\!i\epsilon)^2}{E^2} 
\,,\qquad
\label{Lorentz breaking distance: general}
\end{eqnarray}
where, for gravitational waves oscillating in the $(i,j=x,y)-$plane, we have,
\begin{equation}
{\cal G}_{ij}(u;u') \equiv \begin{cases} 
                                   {\cal G}^\perp_{ij}(u;u'), & {\rm when}\quad i,j=1,2\;;\cr
                                            \delta_{ij},    & {\rm when}\quad  i,j=3,\cdots,D-2\,.     \cr
                     \end{cases}
\label{deformation matrix: extended}
\end{equation}
Eq.~(\ref{Lorentz breaking distance: general}) implies that, transforming from lightcone coordinates 
to Cartesian coordinates is simple, and it amounts to, $(\Delta u\mp i\epsilon)/(2P_v) \rightarrow (\Delta t\mp i\epsilon)/E$, where $E$ is the energy per unit mass.
Equation~(\ref{Lorentz breaking distance: general}) is a remarkable result,
and it states that the only effect planar gravitational waves induce on inertial particles
(moving along free geodesics) as seen by an Unruh-DeWitt detector
is through the prefactor in the Wightman function~(\ref{Wightman function: zero mass}).
Now upon inserting Eq.~(\ref{Wightman functions: general Lor viol solution})
into~(\ref{response function: UdW detector 2}) one obtains,
\begin{eqnarray}
{\cal R}(U,\Delta E)    \!\!&=&\!\! \frac{\hbar m^{D-2}}{(2\pi)^\frac{D}{2}}
\int_{-\infty}^{\infty}\frac{{\rm d}\Delta \tau}
               {[\gamma(u)\gamma(u')]^\frac14\sqrt{{\rm det}[{\cal G}^{ij}(U;\Delta u)]}}
       {\rm e}^{-i\Delta E\Delta \tau}
   \frac{K_{\frac{D-2}{2}}\big(im(\Delta \tau \!-\! i\epsilon)\big)}
   {\big(im(\Delta \tau \!-\! i\epsilon)\,\big)^\frac{D-2}{2}}
\,,\qquad 
\label{transition rate: UdW detector 2}
\end{eqnarray}
where $u=U+\Delta u/2$, $u'=U-\Delta u/2$, and 
we made use of Eq.~(\ref{Lorentz breaking distance: general}),
and we made use of, $m\sqrt{-(\Delta \tau \!-\! i\epsilon)^2}=im(\Delta \tau \!-\! i\epsilon)$. 
Note that in Eq.~(\ref{transition rate: UdW detector 2}) we use the standard $i\epsilon$ prescription
on the Wightman function to describe detector's response rate, which is suitable for problems in which 
the coupling between the detector and the system is time independent, or when it is turned on adiabatically in time. 
In more complicated situations, for example, when the gravitational wave amplitude varies in time, 
a more careful analysis is needed, see {\it e.g.} Ref.~\cite{Louko:2006zv}.

The principal objective of this section is to compute the transition
rate in Eq.~(\ref{transition rate: UdW detector 2}).
 Except for the factor $\big[\sqrt{\gamma(u)\gamma(u')}\,{\rm det}[{\cal G}^{ij}(U;\Delta u)]\big]^{-\frac12}$,
the integrand in
Eq.~(\ref{transition rate: UdW detector 2}) is identical to that for the massive scalar field in Minkowski vacuum.
Since in the on-shell limit (when $u'=u$) this factor equals unity, the effect is purely off-shell, {\it i.e.} 
it occurs due to the off-shell modification of the surface area in which the gravitational waves 
propagate.~\footnote{By the surface in which the gravitational waves propagate we mean the surface orthogonal 
to the direction of propagation, {\it i.e.} it is the $xy$-plane for the waves propagating in the $z$-direction.
The off-shell modification of the surface area for circularly polarized gravitational waves
 is illustrated in figure~3 of Ref.~\cite{vanHaasteren:2022agf}.} 
 The effect is purely off-shell, as it arises as the result of quantum superposition of a single
massive scalar particle (the detector) moving on two distinct trajectories, 
which propagate in a space in which 
the distances in the plane of propagation are contracted (or expanded) by the gravitational backreaction induced by
gravitational waves. Since the transitions arise as a result of quantum superposition of 
different  trajectories, the effect is purely quantum mechanical, and therefore it can be considered as a dynamical 
analogue of the quantum gravitational effect discussed {\it e.g.} in Refs.~\cite{Bose:2017nin,Marshman:2019sne}.

Before we embark on the full calculation, let us firstly consider the 
simpler, massless scalar, case, whose 
Wightman function is obtained by taking the 
limit $m\rightarrow 0$ in Eq.~(\ref{Wightman functions: general Lor viol solution}). The following 
series representation of the Bessel function is handy, 
\begin{equation}
\frac{K_\nu(z)}{z^\nu} = \frac{\Gamma(\nu)\Gamma(1-\nu)}{2^{1+\nu}}
                  \left[\sum_{n=0}^\infty\frac{(z/2)^{2n-2\nu}}{n!\Gamma(n+1-\nu)}
                  -\sum_{n=0}^\infty\frac{(z/2)^{2n}}{n!\Gamma(n+1+\nu)}\right]
\,,\qquad
\label{power series of Bessel function}
\end{equation}
where $z=m\sqrt{\Delta \bar x_{(\pm)}^2}$ and $\nu=(D-2)/2$. In the massless limit only the first term
of the first series in Eq.~(\ref{power series of Bessel function}) contributes, resulting in,
\begin{equation}
i\Delta_0^{(+)}\big(x;x'\big)  
= \frac{\hbar \Gamma(\frac{D-2}{2})}{4\pi^{D/2}
   [\gamma(u)\gamma(u')]^{1/4}\sqrt{{\rm det}[{\cal G}^{ij}(u;u)]}}
    \left(\frac{1}{\Delta \bar x_{(+)}^2(x;x')}\right)^\frac{D-2}{2}
\,,\qquad
\label{Wightman function: zero mass}
\end{equation}
where $\Delta \bar x_{(+)}^2(x;x')=-(\Delta \tau-i\epsilon)^2$ is given in 
Eqs.~(\ref{Lorentz breaking distance: general}),
and whose four dimensional limit is obtained by setting $D=4$.
%

In what follows we evaluate the integral in Eq.~(\ref{transition rate: UdW detector 2}) for two simple cases of 
monochromatic gravitational waves. We shall firstly consider the detector transition rate 
for monochromatic, circularly polarized gravitational waves, and then for 
maximally polarized gravitational waves.
 In this paper we calculate detector's excitation rate, for which 
$\Delta E = E-E_0 >0$. Namely, in realistic situations one expects miniscule detector rates,
and since the excitation rate of the detector in Minkowski vacuum is exactly zero, 
observing any {\it non-vanishing detector's excitation rate} could be interpreted as a signal for 
passing gravitational waves.


\bigskip

\begin{figure}[h!]
\vskip -.1cm
\centerline{\hspace{.in}
\epsfig{file=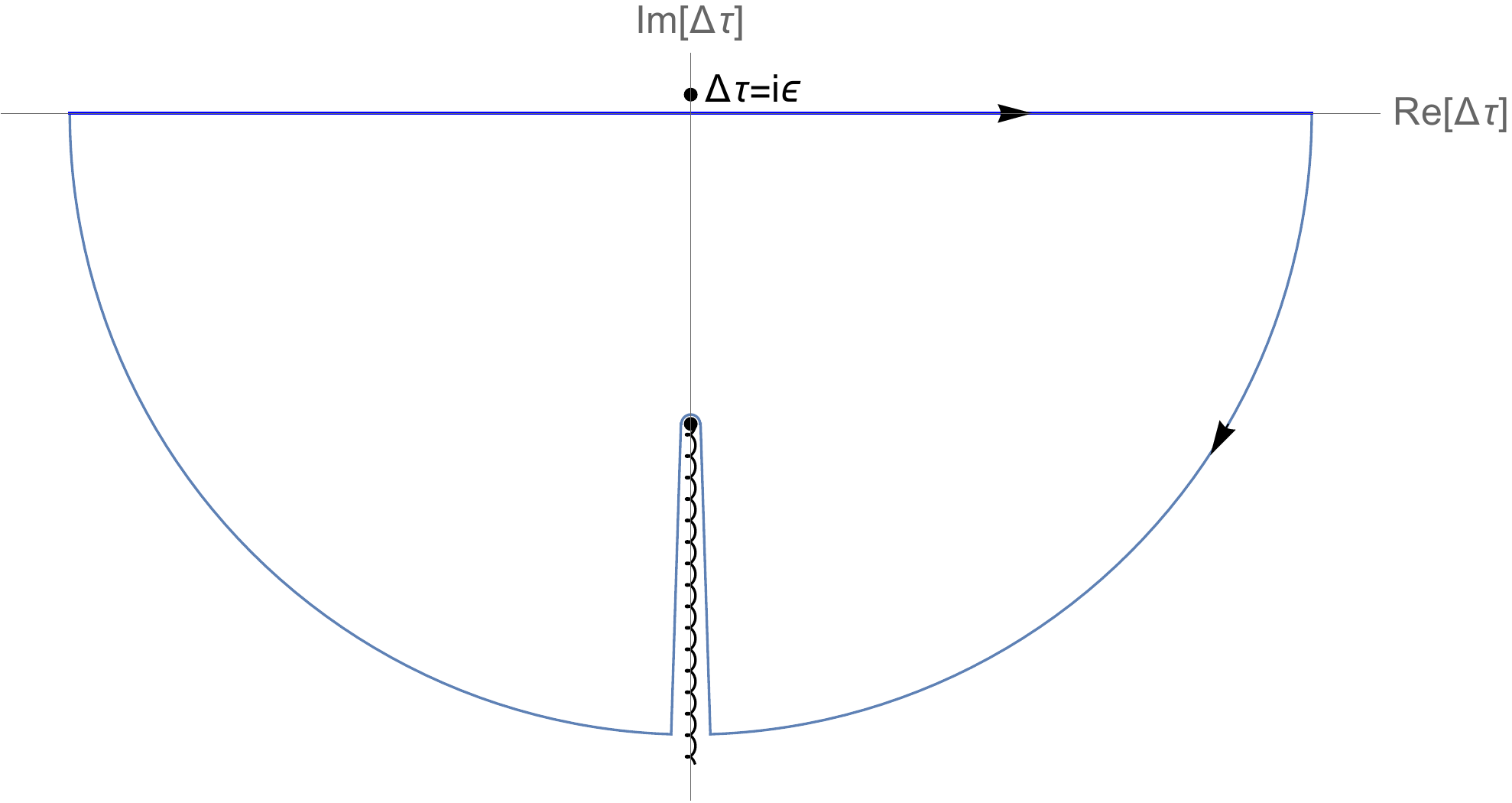, width=4.9in}
}
\vskip -0.1cm
\caption{\small The complex contour used to obtain the detector excitation rate 
in Eq.~(\ref{transition rate: UdW detector 0A}), which shows that the integral 
over the real axis can be replaced by the two sections along the cut in the lower complex $\Delta u-$plane.
}
\label{figure one}
\end{figure}

{\bf Monochromatic circularly polarized gravitational waves.} The deformation 
matrix~(\ref{deformation matrix: extended}) for gravitational waves 
in linear representation~(\ref{metric perturbation}) in the $(xy)-$plane in 
the $\{U,\Delta u\}-$coordinates, 
$\mathbf{\cal G}^\perp_{ij}(U;\Delta u)$, can be inferred from
Eqs.~(\ref{inverse rotation matrix}--\ref{determinant upsilon: linear rep nonpol}),
\begin{eqnarray}
\mathbf{\cal G}^\perp_{ij}(U;\Delta u)
\!\!&=&\!\! \frac{{\rm det}[{\cal G}^\perp_{ij}(\Delta u)]}{\gamma}\left(\!\!\begin{array}{cc}
                 1 \!+\! h\cos(\omega_g U)j_0\left(\frac{\omega_g \Delta u}{2}\right) &
                        h\sin(\omega_g U)j_0\left(\frac{\omega_g \Delta u}{2}\right) \cr
                        h\sin(\omega_g U)j_0\left(\frac{\omega_g \Delta u}{2}\right) &  
                  1 \!-\! h\cos(\omega_g U)j_0\left(\frac{\omega_g \Delta u}{2}\right)\cr
                  \end{array}\!\!\right)
 \nonumber\\
\!\!&=&\!\! \frac{{\cal G}_\perp(\Delta u)}{\gamma}
\left[ \delta^\perp_{ij}+h_{ij}^\perp(U)j_0\left(\frac{\omega_g \Delta u}{2}\right)\right]
\,,\qquad (i,j=1,2)
\,,
\label{inverse rotation matrix 2}\\
\gamma{\cal G}^{-1}_\perp(\Delta u) 
 \!\!&\equiv &\!\! \gamma {\rm det} \left[{\cal G}^{ij}(U;\Delta u)\right]
 =  \frac{1}{\gamma}\bigg[1\!-\!h^2j^2_0\Big(\frac{\omega_g\Delta u}{2}\Big)\bigg]
 \,,\qquad (\gamma = 1-h^2)
\,,\quad
\label{determinant upsilon: linear rep nonpol 2}
\end{eqnarray}
where $U=(u+u')/2$, $\Delta u = u-u'$, 
and $j_0(z) = \sin(z)/z$.~\footnote{In exponential representation used in Ref.~\cite{vanHaasteren:2022agf}, 
in which the spatial part of the metric tensor is, 
$g_{ij}(u) = \left[\exp\left({\mathbf{\tilde h}}(u)\right)\right]_{ij}$,
we have, $\gamma\equiv {\rm det}[g_{ij}(u)]=1$ and 
$\gamma \Upsilon(u;u') = \cosh^2(\tilde h) - \sinh^2(\tilde h)j_0^2\left(\omega_g\Delta u/2\right)$,
such that $\gamma \Upsilon(u;u) =1$. 
 }
For circularly polarized gravitational waves the transition rate 
in Eq.~(\ref{transition rate: UdW detector 2}) simplifies to, 
\begin{eqnarray}
{\cal R}(\Delta E)    \!\!&=&\!\! \frac{\hbar m^{D-2}\sqrt{\gamma}}
               {(2\pi)^\frac{D}{2}}\int_{-\infty}^{\infty}
               \frac{{\rm d}\Delta \tau}{\big[1\!-\!h^2j^2_0\big(P_v\omega_g\Delta\tau\big)\big]^{1/2}} 
{\rm e}^{-i\Delta E\Delta \tau}
   \frac{K_{\frac{D-2}{2}}\big(im(\Delta \tau \!-\! i\epsilon)\big)}
   {\big(im(\Delta \tau \!-\! i\epsilon)\,\big)^\frac{D-2}{2}}
\,,\qquad 
\label{transition rate: UdW detector: nonpol}
\end{eqnarray}
where we made use of, $\Delta u = 2P_v\Delta \tau$. 
Notice that,
for circularly polarized gravitational waves, ${\cal R}(\Delta E)$ does not depend on the average time $U$.
When understood 
as a function of complex $\Delta u$, the integrand in Eq.~(\ref{transition rate: UdW detector: nonpol})
has two square-root cuts along the imaginary axis of complex $\Delta \tau$, 
starting at the roots of the equation, 
\begin{equation}
j_0\big(P_v\omega_g\Delta\tau\big) = \frac{\sinh(\theta)}{\theta} 
          = \frac{1}{h}\gg 1
\,,
\label{cuts of Delta tau: nonpol}
\end{equation}
where $\theta = -i P_v\omega_g\Delta\tau$. For $h\ll 1$ the root $\theta_0(h)$ 
can be approximated by 
the solution of $\theta_0/({\rm e}^{\theta_0}-{\rm e}^{-\theta_0}) = h/2\ll 1
\Rightarrow (-|\theta_0|){\rm e}^{-|\theta_0|}\approx  -h/2$, which can be expressed in terms of 
the Lambert W function (defined by the solution of, 
$w {\rm e}^w =z$),~\footnote{The solutions for the poles, 
$|\theta_0| \approx -\Re[W(-h/2)]$ can be approximated by iterating,
$|\theta_0| = \ln\left(\frac{2}{h}\right)+\ln(|\theta_0|)$, giving
$|\theta_0| = \ln\left(\frac{2}{h}\right)+\ln\big[\ln(\frac{2}{h})+\ln\big(\ln(\frac{2}{h})\cdots\big)\big]$.
The exact solution can be obtained by iterating, 
$|\theta_0| =\ln\left(\frac{2}{h}\right)+\ln\big(\frac{|\theta|}{2}+\sqrt{h^2+\theta_0^2}\big)$. 
When $h\ll 1$, the error in the approximation by the Lambert function decreases as, 
${\cal O}(h^2/\ln^2(h))$,
such that, in the limit when $h\rightarrow 0$, the approximation  by the 
(real part of the) Lambert function becomes exact.
} 
\begin{equation}
|\theta_0(h)| \approx -\Re\big[W\big(\!\!-\frac{h}{2}\big)\big]
\,,
\label{Lambert function approximation}
\end{equation}
such that there are two symmetric solutions.~\footnote{In exponential representation
the approximate roots are given by changing Eq.~(\ref{Lambert function approximation})
to  $|\theta_0(h)| \approx -\Re[W(-\tanh(\tilde h)/2)]$. This means that 
the results in exponential representation are obtained 
from those in linear representation by the replacement, $h \rightarrow \tanh(\tilde h)$.} 
These two roots define the beginning of the square-root cuts, the lower one is 
shown in figure~\ref{figure one}.
The lower cut is responsible for the detector's excitation rate, for which $\Delta E=E-E_0>0$, which is the rate which will be calculated in this paper.~\footnote{If one were interested in 
detector's de-excitation rate stimulated by passing gravitational waves, 
for which $\Delta E<0$,
the complex contour would have to be closed in the upper-half complex $\Delta u$ plane.
The relevant contributions to the rate would then come not only from the cut above the real axis, but also from the pole at $\Delta \tau = i\epsilon$.
Since this pole contributes also in Minkowski space, 
it would be hard to disentangle the pole contributions from those generated by the cut,
and for that reason we do not study these transitions here.
Note that the {\it cut} contribution to the {\it de-excitation rate} can be obtained from the
cut contribution to the excitation
rate simply by exacting the replacement, $\Delta E\rightarrow |\Delta E|$, in the rates.}
The cuts are located at the imaginary $\Delta \tau$-axis, which correspond to spacelike
separations, and they are generated by the coupling between 
the massive scalar and the gravitational waves; the cuts (rather than poles)
arise as a result of the resummed graviton insertions.

One can make use of the Cauchy integral formula to replace the integral in 
Eq.~(\ref{transition rate: UdW detector: nonpol}) by an equivalent integral,
\begin{eqnarray}
{\cal R}(\Delta E)    \!\!&=&\!\! \frac{\hbar m\sqrt{\gamma}}
               {2\pi^2}\int_{\theta_0}^{\infty}
               \frac{{\rm d}\theta}{\big[h^2\sinh^2(\theta)-\theta^2\big]^{1/2}} 
{\rm e}^{-\frac{\Delta E}{P_v\omega_g}\theta}
   K_{1}\big(\frac{m}{P_v\omega_g}\theta\big)
\,,\qquad 
\label{transition rate: UdW detector: nonpol B2}
\end{eqnarray}
%
where we set $D=4$, which is allowed as the integral in Eq.~(\ref{transition rate: UdW detector: nonpol B2})
 is finite in $D=4$, and therefore it does not need to be regularized.
The parameter $\theta_0>0$ in Eq.~(\ref{transition rate: UdW detector: nonpol B2})
is the positive root of Eq.~(\ref{Lambert function approximation}),
we made use of the fact that the integral over the entire contour in figure~\ref{figure one}
vanishes and, in the last step, made the replacement, $\theta\rightarrow -\theta$. 
In the massless limit Eq.~(\ref{transition rate: UdW detector: nonpol B2}) reduces to, 
\begin{eqnarray}
{\cal R}(\Delta E)    \!\!&=&\!\! \frac{\hbar\sqrt{\gamma}P_v\omega_g}
               {2\pi^2}\int_{\theta_0}^{\infty}
               \frac{{\rm d}\theta}{\theta\big[h^2\sinh^2(\theta)-\theta^2\big]^{1/2}} 
{\rm e}^{-\frac{\Delta E}{P_v\omega_g}\theta}
\,.\qquad 
\label{transition rate: UdW detector: nonpol B3}
\end{eqnarray}
%
%
%
The integrals in 
Eqs.~(\ref{transition rate: UdW detector: nonpol B2}--\ref{transition rate: UdW detector: nonpol B3})
are hard, and cannot be evaluated analytically. Let us firstly consider the easier, massless case.

Two analytic approximations can be used for the transition rate in 
Eq.~(\ref{transition rate: UdW detector: nonpol}), an expansion in powers of $h^2$ 
and an expansion around the beginning of the cut, the latter increasing in accuracy in the large 
energy limit, when $\Delta E \gg P_v\omega_g$. 
The first approximation amounts to setting $D=4$
and expanding~(\ref{transition rate: UdW detector: nonpol})
in powers of $h^2$. This replaces the cut contribution by a sum over the poles 
at $\Delta u = i\epsilon$ of the order $2n+2$,
\begin{eqnarray}
{\cal R}(\Delta E)    \!\!&=&\!\! \frac{\hbar m^{2}\sqrt{\gamma}}{4\pi^2}
    \frac{1}{\sqrt{\pi}}\sum_{n=1}^\infty \Gamma\big(n\!+\!\frac12\big) \frac{h^{2n}}{n!}\!
    \int_{-\infty}^{\infty}\!
               \frac{{\rm d}\Delta \tau}{(P_v\omega_g\Delta\tau)^{2n}} 
               \sin^{2n}(P_v\omega_g\Delta\tau)
{\rm e}^{-i\Delta E\Delta \tau}
   \frac{K_1\big(im(\Delta \tau \!-\! i\epsilon)\big)}
   {im(\Delta \tau \!-\! i\epsilon)}
,\qquad \;\;\;
\label{transition rate: UdW detector: nonpol power series}
\end{eqnarray}
which can be evaluated by making use of the Cauchy integral formula.
The integration of the $n$th term in the sum does not vanish provided $\Delta E - 2nP_v\omega_g<0$.
In the massless limit the last term in Eq.~(\ref{transition rate: UdW detector: nonpol})
simplifies to, $-1/[m(\Delta\tau-i\epsilon)]^2$, such that the integral evaluates to, 
\begin{eqnarray}
{\cal R}(\Delta E)    \!\!&=&\!\! \frac{\hbar\sqrt{\gamma}P_v\omega_g}{2\pi^2}
                    \Bigg\{\frac{\pi}{6}h^{2} \!\left(\!1\!-\!\frac{\Delta E}{2P_v\omega_g}\right)^3\!
                     \Theta(2P_v\omega_g\!-\!\Delta E)
\label{transition rate: UdW detector: nonpol n=1 and 2}\\
 \!\!&+&\!\!\frac{\pi}{5}h^4 \left[-\frac{1}{8}\left(\!1\!-\!\frac{\Delta E}{2P_v\omega_g}\right)^3\!
                     \Theta(2P_v\omega_g\!-\!\Delta E)
                     \!+\!\left(\!1\!-\!\frac{\Delta E}{4P_v\omega_g}\right)^3\!
                     \Theta(4P_v\omega_g\!-\!\Delta E)\right] \!+\!{\cal O}(h^6)
          \Bigg\}
\,,\qquad 
\nonumber
\end{eqnarray}
where we made use of, $\sin^2(z) = \frac12 - \frac14 \left({\rm e}^{2iz} +{\rm e}^{-2iz}\right)$,
and we have assumed that, $\Delta E>0$ and $P_v>0$.

\begin{figure}[h!]
\vskip -.4cm
\centerline{\hspace{.in}
\epsfig{file=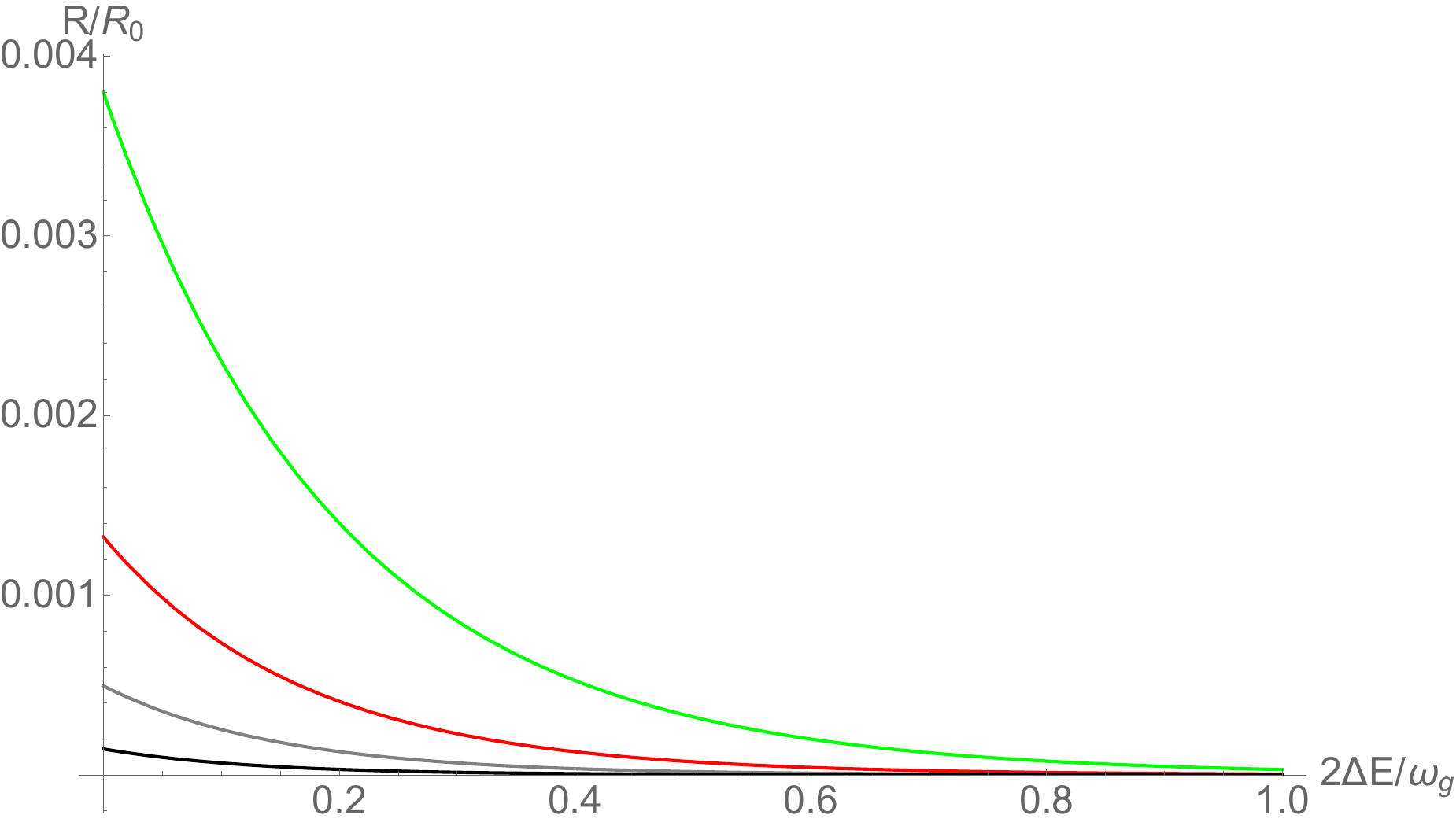, width=3.3in}
\hskip 1cm
\epsfig{file=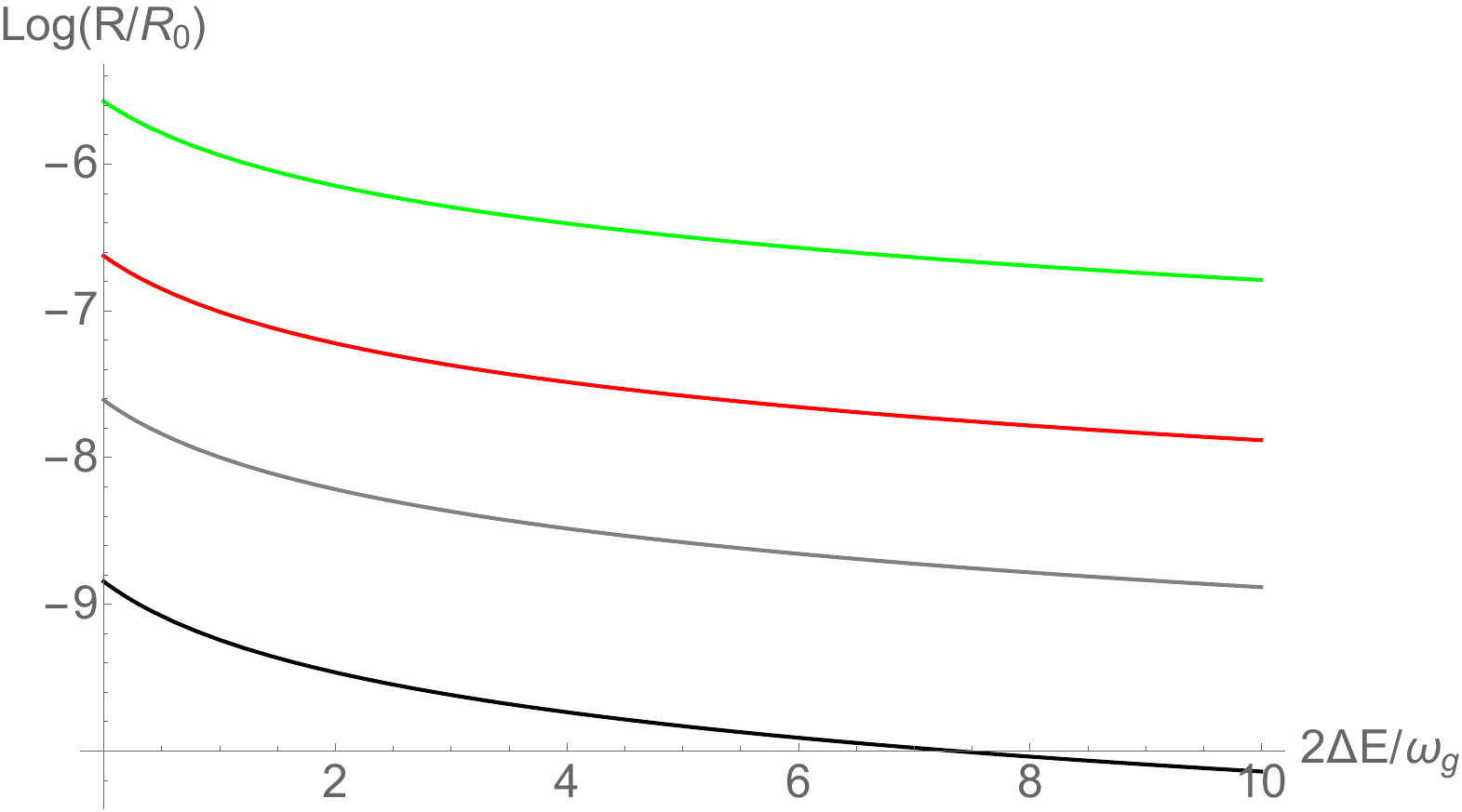, width=3.in}
}
\vskip -0.1cm
\caption{\small {\it Left panel:} The transition rate ${\cal R}$ 
in Eq.~(\ref{transition rate: UdW detector: nonpol B3}) for a massless scalar 
as a function of $\Delta E/(P_v \omega_g)$
in units of ${\cal R}_0=\hbar \sqrt{1-h^2} P_v\omega_g/(2\pi ^2)$ 
as measured by the Unruh-DeWitt detector. Different curves are for different values of 
the gravitational strain (from top down): $h=0.1$ (green),  $h=0.05$ (red), $h=0.025$ (gray), and $h=0.01$ (black).
{\it Right panel:} The same diagram for $\ln({\cal R}/{\cal R}_0)+\Delta E \theta_0/(P_v\omega_g)$, where 
$\theta_0>0$ is the imaginary pole at the beginning of the square root cut
in Eq.~(\ref{cuts of Delta tau: nonpol}).
}
\label{figure two}
\end{figure}

\begin{figure}[h!]
\vskip -.2cm
\centerline{\hspace{.in}
\epsfig{file=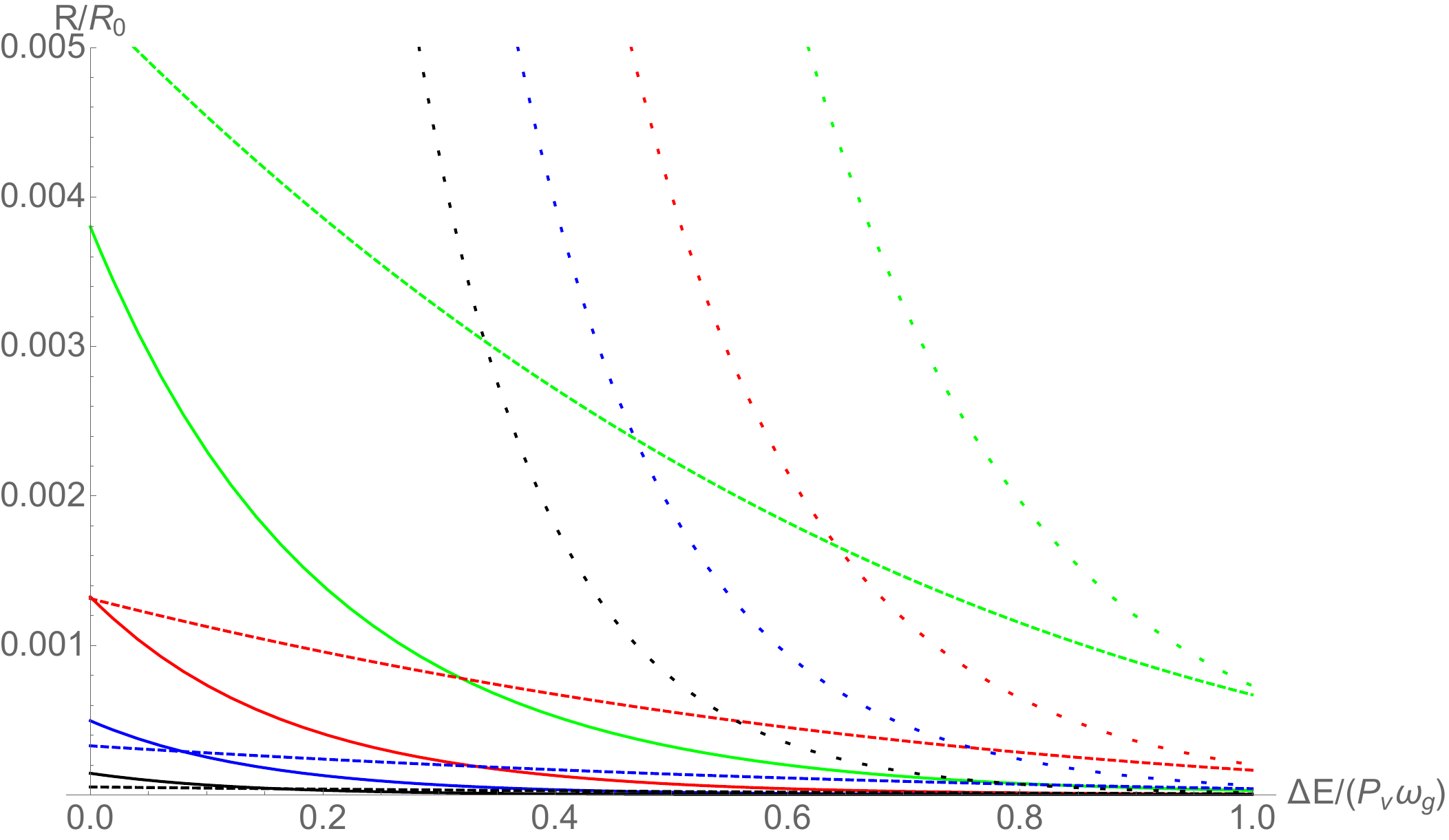, width=3.6in}
\hskip 0.1cm
\epsfig{file=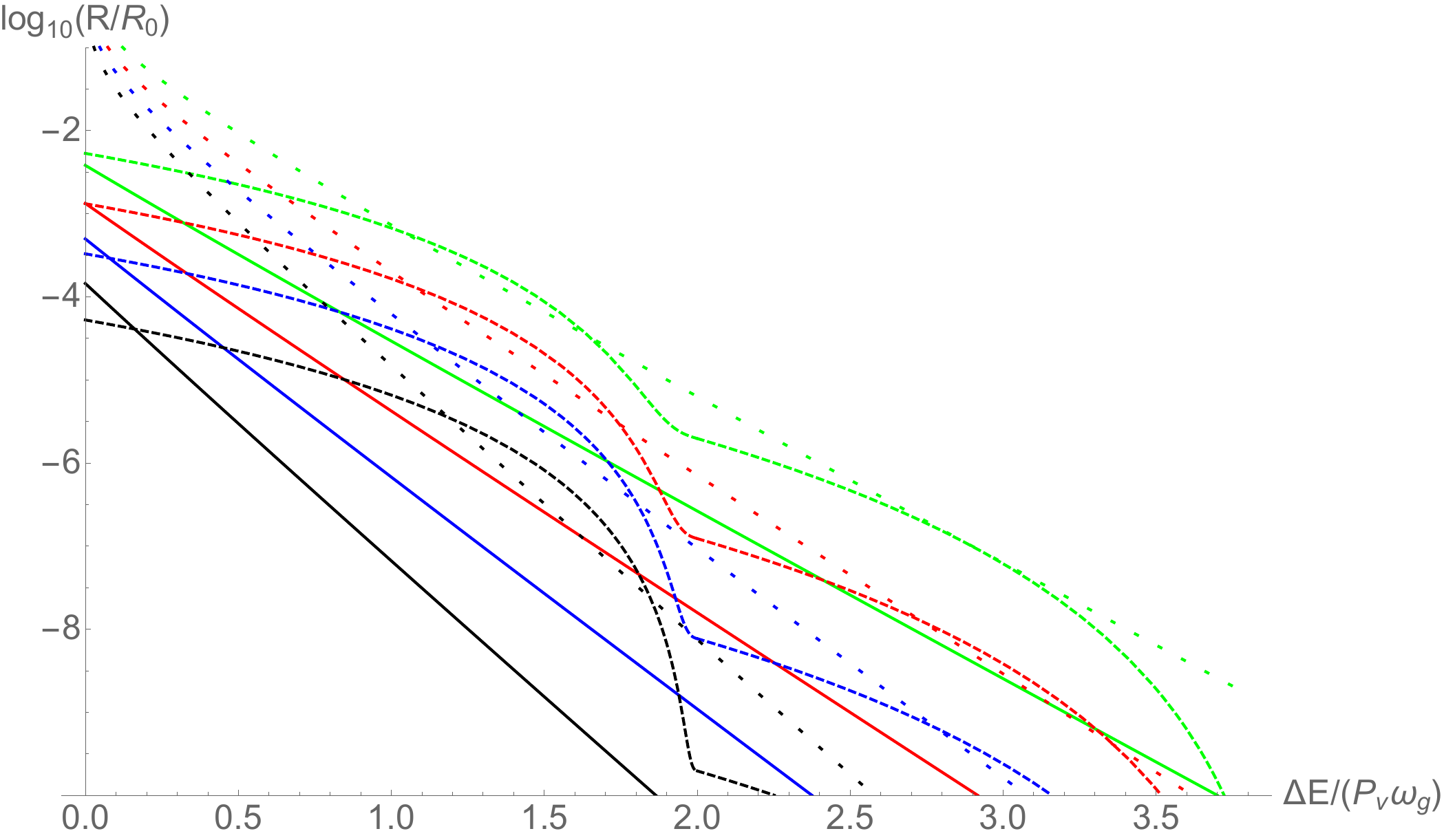, width=3.6in}
}
\vskip -0.1cm
\caption{\small {\it Left panel:} The detector transition rate ${\cal R}$ for a massless scalar 
as a function of $\Delta E/(P_v \omega_g)$
in units of ${\cal R}_0=\hbar \sqrt{1-h^2} P_v\omega_g/(2\pi ^2)$ 
as measured by the Unruh-DeWitt detector. Different curves are for different values of 
the gravitational strain: $h=0.1$ (green),  $h=0.05$ (red), $h=0.025$ (blue), and $h=0.01$ (black).
Solid lines show numerical results, dashed lines are approximate curves
in Eq.~(\ref{transition rate: UdW detector: nonpol n=1 and 2}), 
obtained by expanding the integrand in powers of $h^2$, and dotted lines 
represent the function in Eq.~(\ref{transition rate: UdW detector: nonpol: analytic 2}), 
obtained by expanding the integrand around the beginning of the cut.
{\it Right panel:} The same diagram for $\log_{10}({\cal R}/{\cal R}_0)$.
}
\label{figure three}
\end{figure}

The second analytic approximation can be obtained by expanding the integrand in 
Eq.~(\ref{transition rate: UdW detector: nonpol B3}) around the beginning of the cut. The result is, 
\begin{eqnarray}
{\cal R}(\Delta E)    \!\!&\approx&\!\! \frac{\hbar\sqrt{\gamma}P_v\omega_g}
               {2\pi^2}\sqrt{\frac{P_v\omega_g}{\Delta E}}\left(\frac{h}{2}\right)^\frac{\Delta E}{P_v\omega_g}
               \frac{1}{\left[\ln\left(\frac{2}{h}\right)+\ln\left(\ln\left(\frac{2}{h}\right)\right)\right]^{2+\frac{\Delta E}{P_v\omega_g}}}\left[1+{\cal O}\left(\frac{P_v\omega_g}{\Delta E}\right)\right]
\,.\qquad 
\label{transition rate: UdW detector: nonpol: analytic 2}
\end{eqnarray}
One can evaluate Eq.~(\ref{transition rate: UdW detector: nonpol B3}) numerically, 
and the results are shown in figure~\ref{figure two}. 
which probe different spatial gravitational potential induced by the gravitational backreaction from gravitational waves, 
and it is therefore purely quantum.The effect originates from the quantum interference between propagation on off-shell detector's trajectories
which probe different spatial gravitational potential induced by the gravitational backreaction from gravitational waves, 
and it is therefore purely quantum.
In figure~\ref{figure three} we compare the numerical results for the transition rate
with two analytical approximations.
The first approximation is obtained by expanding in powers of $h^2$ 
(dashed lines in figure~\ref{figure three}) and the second by expanding around the beginning of the cut
(dotted lines in figure~\ref{figure three}). Both approximations are reasonable, but neither is accurate.
The second approximation captures correctly the large $\Delta E$ behavior,
${\cal R}(\Delta E) \propto {\rm exp}\left[-\frac{\Delta E}{P_v\omega_g}\theta_0\right]$,
meaning that the transition rate is exponentially 
suppressed with growing $\left[\frac{\Delta E}{P_v\omega_g}\theta_0\right]$, but the analytical
estimate poorly approximates the constant in front of the exponential. 
Due to the complicated dependence of $\theta_0(h)$ on
$h$ in Eq.~(\ref{Lambert function approximation}), the dependence of the transition rate 
on $h$ is nonanalytic, which explains why expanding in powers of $h^2$ performs relatively poorly.
This kind of nonanalytic behavior is hard to guess, and impossible to obtain without 
knowing the scalar Wightman function from Ref.~\cite{vanHaasteren:2022agf}, 
which resums the gravitational wave insertions. 

The effect originates from the quantum interference between propagation on off-shell detector's trajectories
which probe different spatial gravitational potential induced by the gravitational backreaction from gravitational waves, 
and it is therefore purely quantum.

\begin{figure}[h!]
\vskip -.2cm
\centerline{\hspace{.in}
\epsfig{file=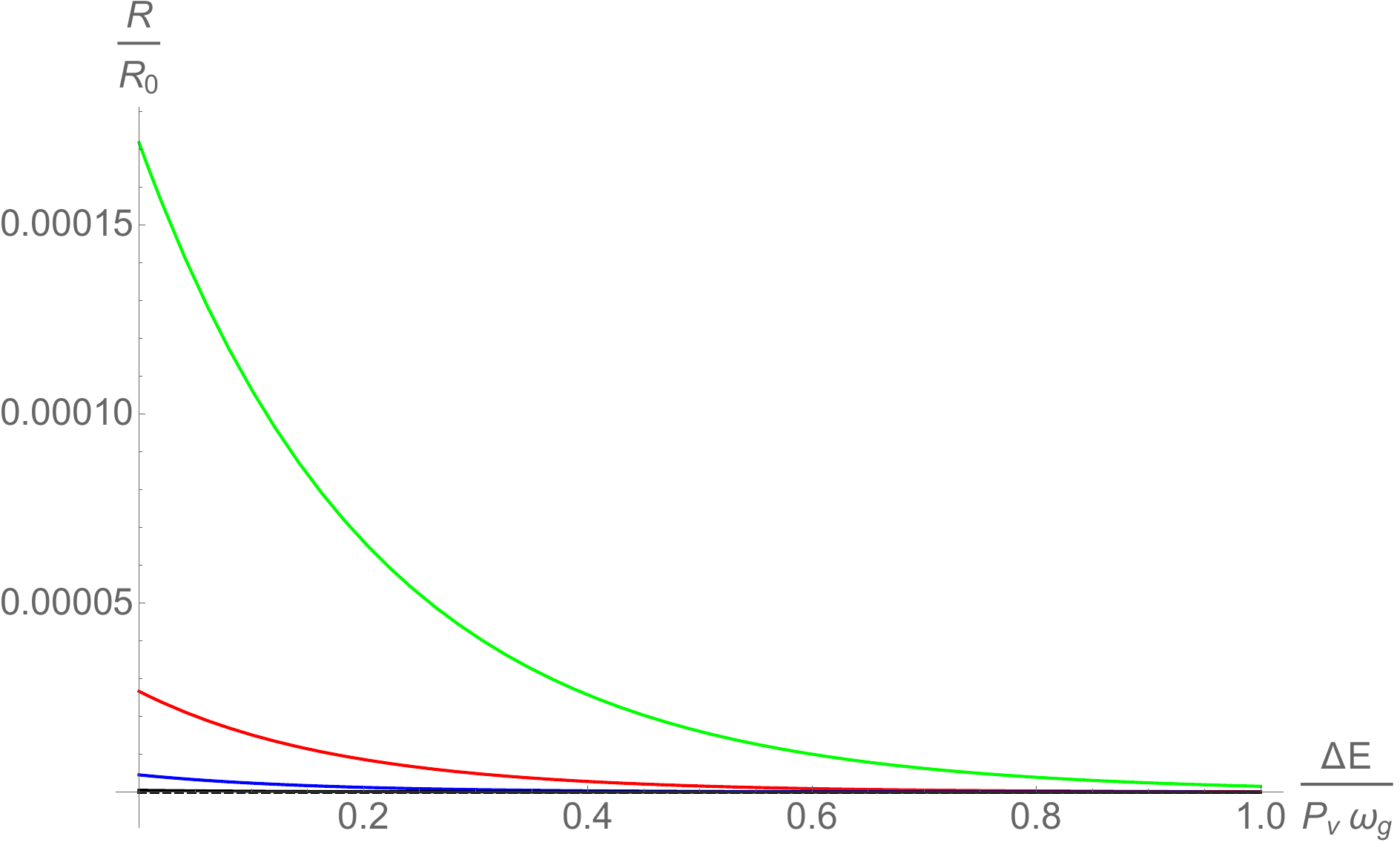, width=3.8in}
\hskip -0.4cm
\epsfig{file=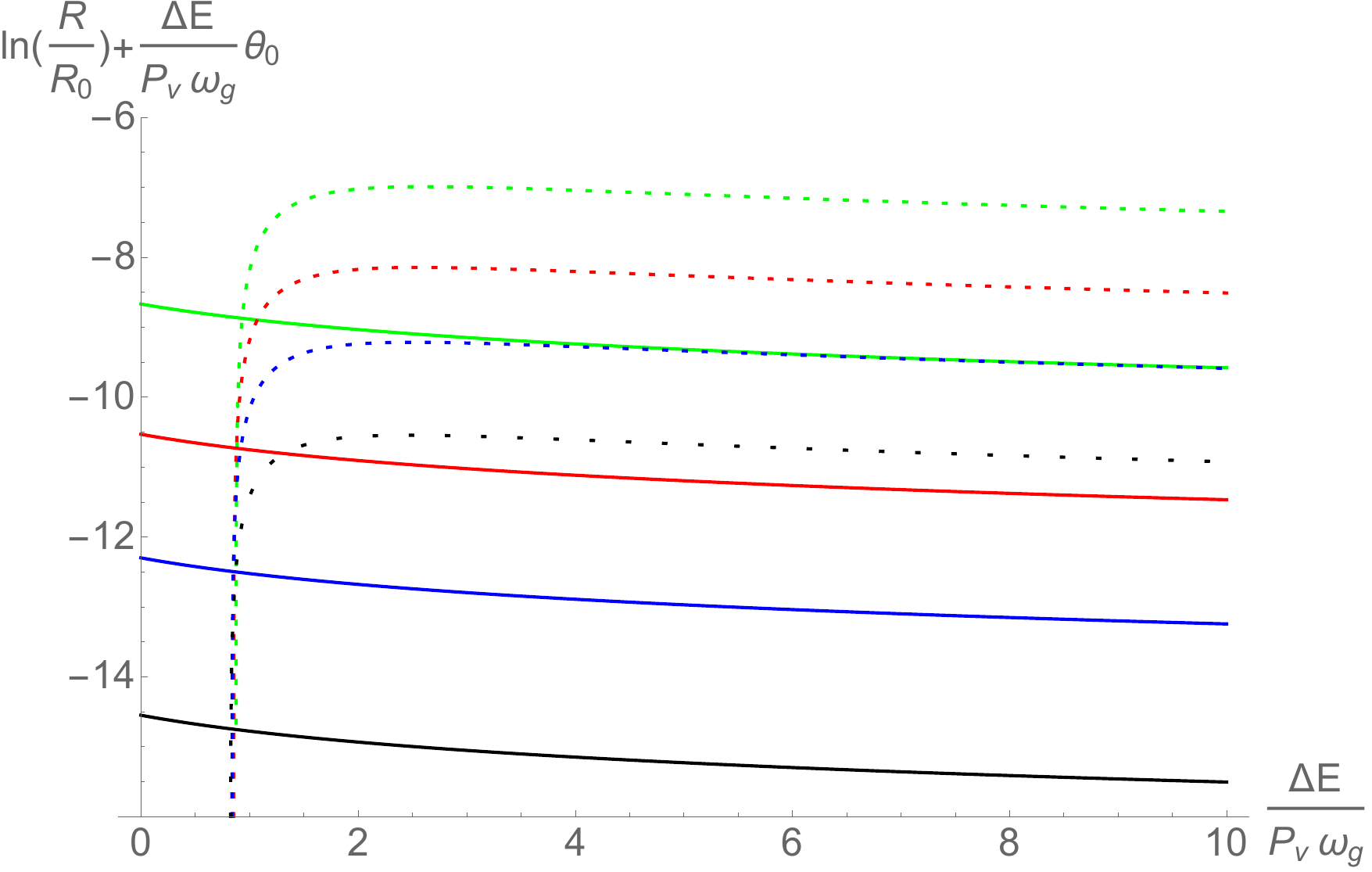, width=3.6in}
}
\vskip -0.1cm
\caption{\small {\it Left panel:} The detector transition rate ${\cal R}(\Delta E)$ for a massive scalar 
with $m=P_v\omega_g$ 
as a function of $\Delta E/(P_v \omega_g)$
in units of ${\cal R}_0=\hbar \sqrt{1-h^2} P_v\omega_g/(2\pi ^2)$ 
as measured by the Unruh-DeWitt detector. Different curves are for different values of 
the gravitational strain: $h=0.1$ (green),  $h=0.05$ (red), $h=0.025$ (blue), and $h=0.01$ (black).
{\it Right panel:} The same diagram for $\ln({\cal R}/{\cal R}_0) +(\Delta E \theta_0)/(P_v\omega_g)$.
}
\label{figure four}
\end{figure}

\begin{figure}[h!]
\vskip -.2cm
\centerline{\hspace{.in}
\epsfig{file=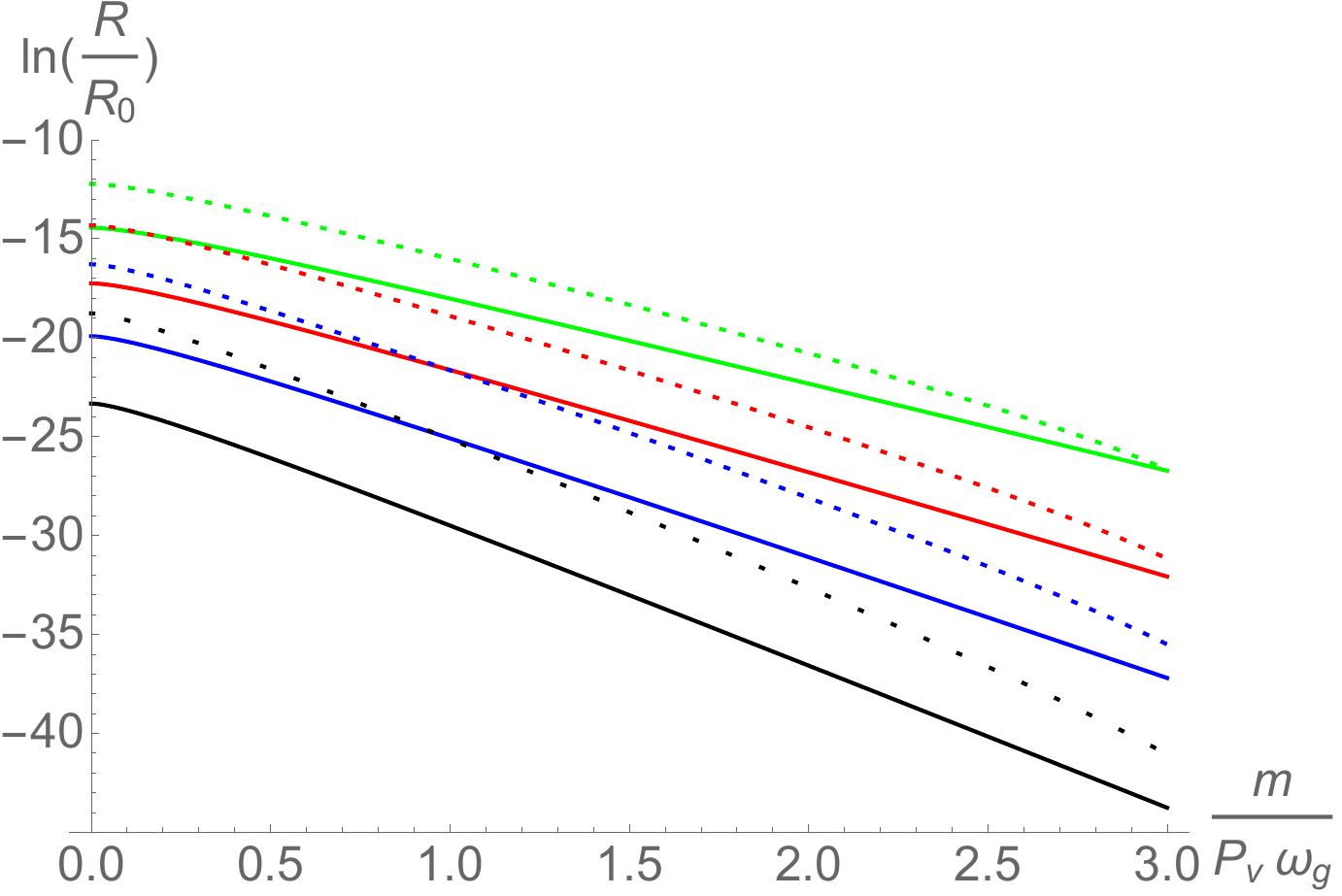, width=3.5in}
\hskip -0.2cm
\epsfig{file=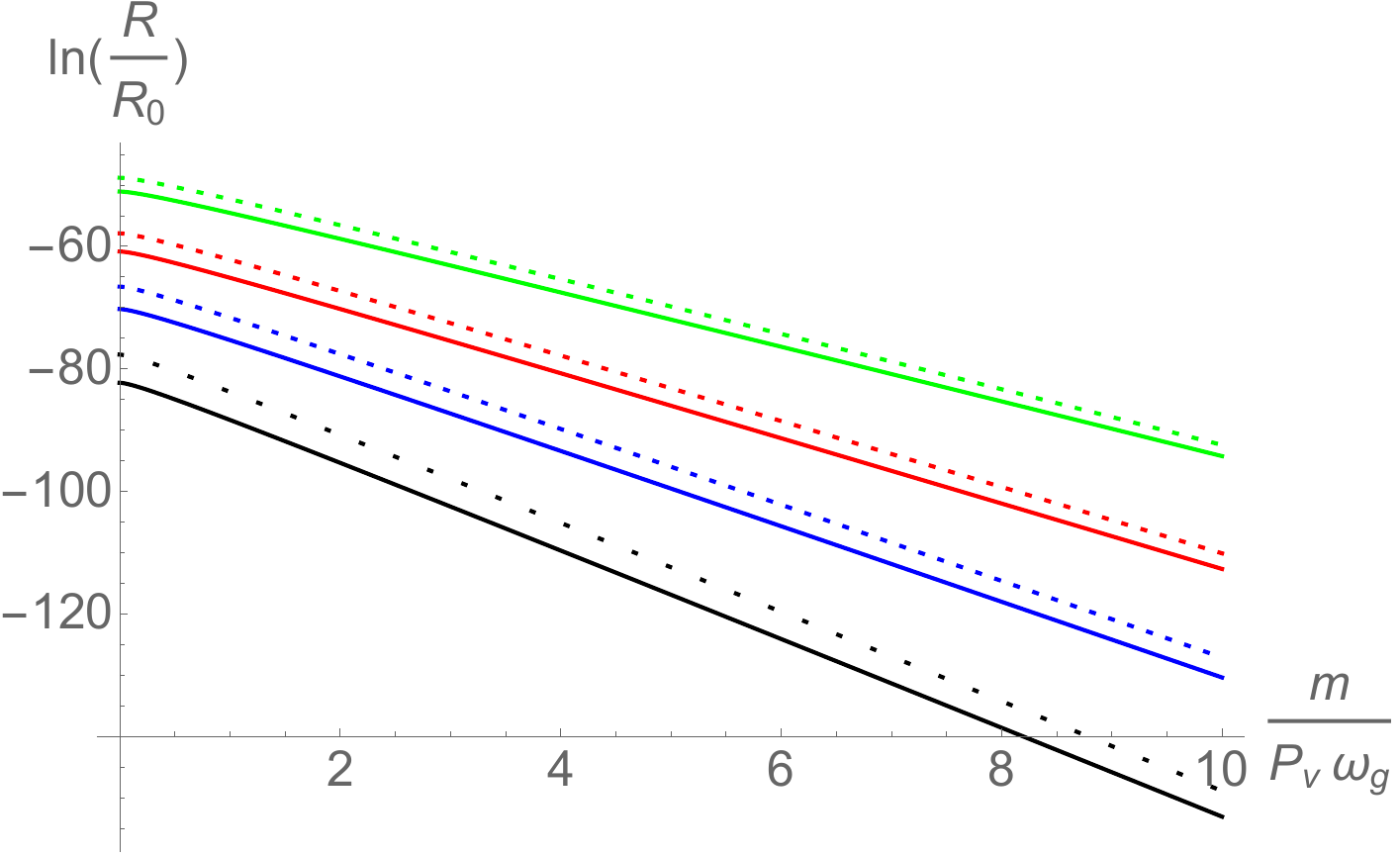, width=3.8in}
}
\vskip -0.1cm
\caption{\small {\it Left panel:} The detector transition rate ${\cal R}$ for a massive scalar 
with $\Delta E/(P_v \omega_g)=2$ 
as a function of $m/(P_v\omega_g)$ 
in units of ${\cal R}_0=\hbar \sqrt{1-h^2} P_v\omega_g/(2\pi ^2)$ 
as measured by the Unruh-DeWitt detector. Different curves are for different values of 
the gravitational strain: $h=0.1$ (green),  $h=0.05$ (red), $h=0.025$ (blue), and $h=0.01$ (black).
{\it Right panel:} The same but with with $\Delta E/(P_v \omega_g)=10$. 
The solid curves are numerical results, and the short dashed curves are 
the approximation in Eq.~(\ref{transition rate: UdW detector: massive nonpol cut}).
}
\label{figure five}
\end{figure}

Switching on the scalar mass suppresses the transition rate further, and the results 
obtained by numerically integrating~(\ref{transition rate: UdW detector: nonpol B2})
for $m=P_v\omega_g$ shown in figure~\ref{figure four} are significantly suppressed when compared with
those for the massless scalar in figures~\ref{figure two} and~\ref{figure three}.
Just as in the massless case, at large $\Delta E$, the results decay exponentially as,
${\cal R}\propto \exp\left[-\frac{\Delta E}{P_v\omega_g}\theta_0\right]$, and in the limit of a large mass,
$m\theta_0\gg P_v\omega_g$, 
there is an additional exponential suppression, 
${\cal R}\propto \exp\left[-\frac{(\Delta E+m)}{P_v\omega_g}\theta_0\right]$. To see that, let us 
approximately evaluate the integral in Eq.~(\ref{transition rate: UdW detector: nonpol B2}) 
by expanding around the beginning of the cut at $\theta=\theta_0(h)$, 
\begin{eqnarray}
{\cal R}(\Delta E)    \!\!&= &\!\! \frac{\hbar\sqrt{\gamma}m}{2\pi^2}
                    \sqrt{\frac{\pi P_v\omega_g}{2\Delta E\theta_0(\theta_0-1)}}
                                             K_1\left(\frac{m\theta_0}{P_v\omega_g}\right)
                       {\rm e}^{-\frac{\Delta E\theta_0}{P_v\omega_g}}
  \nonumber\\
\!\!&\times &\!\! 
                    \Bigg\{1-\frac{P_v\omega_g}{2\Delta E} 
                       \Bigg[\frac{2\theta_0^2-1}{4\theta_0(\theta_0-1)}
         +\frac{m}{P_v\omega_g}\frac{K_2\left(\frac{m\theta_0}{P_v\omega_g}\right)}
                                                    {K_1\left(\frac{m\theta_0}{P_v\omega_g}\right)} - \frac{1}{\theta_0}
            \Bigg]+{\cal O}(\Delta E^{-2})\Bigg\}
\label{transition rate: UdW detector: massive nonpol cut}
\,,\qquad 
\end{eqnarray}
which applies when $\Delta E\theta_0/(P_v\omega_g)\gg1 $, and where we dropped the term
$h^2/(2\theta_0)$ in the expansion,
$\sqrt{\theta_0^2+h^2} \simeq \theta_0 + {\cal O}(h^2/\theta_0)$. This amounts to 
approximating the position of the cut, $\theta=\theta_0(h)$, by the 
Lambert function in Eq.~(\ref{Lambert function approximation}).  
We shall not attempt to evaluate 
the integrals in Eq.~(\ref{transition rate: UdW detector: nonpol power series})
for the massive case, as that would require not only to account for the
contributions of the poles at $\Delta u = i\epsilon$, 
but also for
the contribution of the logarithmic cut 
of the modified Bessel function, 
$im(\Delta u-i\epsilon)K_1\big(im(\Delta u-i\epsilon)\big)$, which extends from $\Delta u = 0$ to $0+i\infty$. 
Finally, in figure~\ref{figure five} we show the same transition rate as in figure~\ref{figure four},
but now as a function of the mass for a fixed $\Delta E = 2 P_v\omega_g$ (left panel) and 
$\Delta E = 10 P_v\omega_g$ (right panel). The figure shows that the  transition rate is exponentially 
suppressed by the mass as, ${\cal R}\sim {\rm e}^{-m\theta_0(h)/(P_v\omega_g)}$, which can be also
inferred from the asymptotic expansion of the Bessel function 
in Eq.~(\ref{transition rate: UdW detector: massive nonpol cut}),
$K_1\left(\frac{m\theta_0}{P_v\omega_g}\right)
          \sim \sqrt{\frac{\pi P_v\omega_g}{2m\theta_0 }}\exp\left(-\frac{m\theta_0}{P_v\omega_g}\right)$.
%


\bigskip

{\bf Monochromatic elliptically polarized gravitational waves.} Here we shall consider 
general elliptically polarized gravitational
waves, whose analysis is much trickier, as the interaction with the detector 
depends on the average time variable, $U=(u+u')/2$. Let us begin our analysis by noting that 
Eq.~(\ref{deformation matrix: momentum space}) can be recast as, 
\begin{equation}
{\cal G}^{ij}(U,\Delta u) =\sum_{n=0}^\infty \left[\frac{\partial^{2n}}{\partial U^{2n}}g^{ij}(U)\right]
                      \frac{\left({\Delta u}/{2}\right)^{2n}}{(2n+1)!}
\,,\qquad
\label{deformation matrix: momentum space: pol}
\end{equation}
where $\Delta u = u\!-\!u'$, and $ g^{ij}(U)$ denotes the inverse of $g_{ij}(U)$,
which for generally polarized waves (fluctuating in a 2-dimensional plane) takes the form
({\it cf.} Eq.~(\ref{deformation matrix: momentum space 2})), 
\begin{eqnarray}
g^{ij}(U) \!\!&=&\!\! \frac{1}{1-h_+^2c_+^2-h_\times^2 c_\times^2}
               \left(\!\!\begin{array}{cc} 
                       1 - h_+c_+ & -h_\times c_\times\cr 
                        -h_\times c_\times & 1 + h_+c_+  \cr 
                       \end{array}
                \!\!\right)
\,,\qquad
\label{deformation matrix: momentum space: pol 2}
\end{eqnarray}
where 
$c_+(U)=\cos(\omega_gU+\psi/2)$, $c_\times(U) =\cos(\omega_gU-\psi/2)$,
where $\psi$ is an arbitrary phase (in the case of nonpolarized gravitational waves, $\psi=\pi/2$).
Evaluating~(\ref{deformation matrix: momentum space: pol}) for 
$g^{ij}$ in Eq.~(\ref{deformation matrix: momentum space: pol 2}) 
is a formidable task, and we shall content ourselves by evaluating 
${\cal G}^{ij}(u;u')$ to the quadratic order in $h_+$ and $h_\times$,
\begin{eqnarray}
{\cal G}^{ij}(U,\Delta u)  \!\!&=&\!\! \left(\!\!\begin{array}{cc} 
                       1 & 0 \cr 
                       0 & 1 \cr
                       \end{array}
                \!\!\right)  \left[1+\frac{h_+^2}{2}\left(1+(2c_+^2-1)j_0(\omega_g\Delta u)\right)
                              +\frac{h_\times^2}{2}\left(1+(2c_\times^2-1)j_0(\omega_g\Delta u)\right)\right]
\nonumber\\
 \!\!&&\!\! +\,\left(\!\!\begin{array}{cc} 
                       - h_+c_+ & -h_\times c_\times\cr 
                        -h_\times c_\times & h_+c_+  \cr 
                       \end{array}
                \!\!\right)j_0\left(\frac{\omega_g\Delta u}{2}\right) 
                +{\cal O}(h_+^2h_\times,h_+h_\times^2)
\,,\qquad
\label{deformation matrix: momentum space: pol 3}
\end{eqnarray}
whose determinant equals, 
\begin{eqnarray}
{\cal G}^{-1}(U;\Delta u)  \!\!&=&\!\! 1 - h_+^2 \left[c_+^2 j_0^2\left(\frac{\omega_g\Delta u}{2}\right) 
               -1-(2c_+^2-1)j_0(\omega_g\Delta u)
  \right]
  \qquad\quad
\nonumber\\
 \!\!&-&\!\!h_\times^2 \left[c_\times^2 j_0^2\left(\frac{\omega_g\Delta u}{2}\right) 
    - 1-(2c_\times^2-1)j_0(\omega_g\Delta u)
\right]
                +{\cal O}(h_+^2h_\times,h_+h_\times^2)
\,.\qquad
\label{deformation matrix: momentum space: pol 3}
\end{eqnarray}
In the circularly polarized case, in which $\psi=\pi/2$ and $h_+=h_\times =h$, this reduces to, 
${\cal G}^{-1}(U;\Delta u) \rightarrow 1 - h^2 \left[j_0^2\left(\frac{\omega_g\Delta u}{2}\right) +2\right]$,
which agrees with Eq.~(\ref{determinant upsilon: linear rep nonpol 2}) when one recalls that, 
$1/\gamma^2 = 1+2h^2+{\cal O}(h^4)$. Upon introducing 
$\tilde \gamma(h_+,h_\times)=1-\frac{h_+^2+h_\times^2}{2}$, 
Eq.~(\ref{deformation matrix: momentum space: pol 3}) can be recast as, 
\begin{eqnarray}
{\cal G}^{-1}(U;\Delta u)  \!\!&=&\!\!
\frac{1}{\tilde \gamma^2}\left\{ 1 - \left[\big(h_+^2 c_+^2+h_\times^2c_\times^2\big)
                                j_0^2\left(\frac{\omega_g\Delta u}{2}\right) 
        -\big(h_+^2(2c_+^2-1)+h_\times^2(2c_\times^2-1)\big)j_0(\omega_g\Delta u)
  \right]
  \right\}  
  \qquad\;
\nonumber\\
 \!\!&&\!\! +\,{\cal O}(h_+^2h_\times,h_+h_\times^2)
\,.\qquad
\label{deformation matrix: momentum space: pol 3B}
\end{eqnarray}
To complete the analysis, we also need,
\begin{equation}
\sqrt{\gamma(u)\gamma(u')}  = \tilde \gamma
                      - \frac12\left(h_+^2 (2c_+^2-1)+h_\times^2 (2c_\times^2-1)\right)
                                     \cos(\omega_g\Delta u)
+\,{\cal O}(h_+^2h_\times,h_+h_\times^2)
\,,\qquad
\label{deformation matrix: momentum space: pol 4}
\end{equation}
which, when multiplied with~(\ref{deformation matrix: momentum space: pol 3B}), gives,
\begin{eqnarray}
\sqrt{\gamma(u)\gamma(u')}{\cal G}^{-1}(U;\Delta u)  \!\!&=&\!\!
\frac{1}{\tilde \gamma}\bigg\{ 1 - \bigg[\big(h_+^2 c_+^2+h_\times^2c_\times^2\big)
                                j_0^2\left(\frac{\omega_g\Delta u}{2}\right) 
\label{deformation matrix: momentum space: pol 5}
\\
 \!\!&&\!\! \hskip -2cm
        -\,\big(h_+^2(2c_+^2-1)+h_\times^2(2c_\times^2-1)\big)
         \Big(j_0(\omega_g\Delta u)-\frac12\cos(\omega_g\Delta u)\Big)
  \bigg]
  \bigg\}  
+\,{\cal O}(h_+^2h_\times,h_+h_\times^2)
\nonumber
\,.\qquad
\end{eqnarray}
This product appears in the propagator in Eq.~(\ref{transition rate: UdW detector 2}),
and generates square-root cuts~\footnote{Our analysis is accurate at the order $h^2$, and thus not exact. Therefore, 
one should be aware of the possibility that polarized gravitational wave 
may generate more baroque cuts in the complex $\Delta u$ plane, for an illustration see 
Appendix~B.}, just as in the circularly polarized waves in Eq.~(\ref{transition rate: UdW detector: nonpol}).
The integral can be evaluated by contour integration, with the contour showed in figure~\ref{figure one},
resulting in the cut contribution ({\it cf.} Eq.~(\ref{transition rate: UdW detector: nonpol B2})),
\begin{eqnarray}
{\cal R}(U,\Delta E)    \!\!&=&\!\! \frac{\hbar m\sqrt{\tilde \gamma}}
               {2\pi^2}\int_{\theta_0}^{\infty}
               \frac{{\rm d}\theta}{\big[H^2(U,\theta)-\theta^2\big]^{1/2}} 
{\rm e}^{-\frac{\Delta E}{P_v\omega_g}\theta}
   K_{1}\big(\frac{m}{P_v\omega_g}\theta\big)
\,,\qquad 
\label{transition rate: UdW detector: pol B2}
\end{eqnarray}
where $h^2(U)=h_+^2 c_+^2(U)+h_\times^2c_\times^2(U)$,
\begin{equation}
H^2(U,\theta) = h^2(U)\sinh^2(\theta) 
+ \Big[h^2(U) -\frac{h_+^2+h_\times^2}{2}\Big]\Big[\theta^2\cosh(2\theta)-\theta\sinh(2\theta)\Big]
\,,\qquad\;
\label{function H}
\end{equation}
and $\theta_0>0$ denotes the beginning of the cut defined by, $H^2(U,\theta_0)=\theta_0^2$. 
Next we insert,
\begin{equation}
h^2(U) = \frac{h_+^2\!+\!h_\times^2}{2}+\frac{h_+^2\!+\!h_\times^2}{2}\cos(2\omega_gU)c_\psi
               - \frac{h_+^2\!-\!h_\times^2}{2}\sin(2\omega_gU)s_\psi
,\quad \big(c_\psi=\cos(\psi),\;s_\psi=\sin(\psi)\big),
\;
\label{function h2(U)}
\end{equation}
into Eq.~(\ref{function H}) to obtain, 
\begin{eqnarray}
&&\bigg[\frac{h_+^2\!+\!h_\times^2}{2}c_\psi\cos(2\omega_gU)-\frac{h_+^2\!-\!h_\times^2}{2}s_\psi\sin(2\omega_gU)\bigg]
\Big[\sinh^2(\theta_0)+\theta_0^2\cosh(2\theta_0)-\theta_0\sinh(2\theta_0)\Big]
\nonumber\\
&& \hskip 10cm +\, \frac{h_+^2+h_\times^2}{2}\sinh^2(\theta_0)
        -\theta_0^2=0
\,.\qquad
\label{function H2}
\end{eqnarray}
Because of the $U$ dependence, Eq.~(\ref{function H2}) may or may not
have a solution, meaning that the cuts exist only when~(\ref{function H2}) can be solved for some
real $U$. To simplify our analysis, recall that we are interested in the limit when, 
$h_+,h_\times\ll 1$, in which case in most of the 
parameter space $\theta_0\gg1$, and one can approximate Eq.~(\ref{function H2})
by keeping the leading order terms $\propto {\rm e}^{2\theta_0}$ only.~\footnote{The same
approximation was shown to work extremely well when we analyzed the circularly polarized case.} 
Multiplying Eq.~(\ref{function H2}) by 
$2{\rm e}^{-2\theta_0}/\cos^2(\omega_g U)=2{\rm e}^{-2\theta_0}(1+t^2)$ and neglecting the terms 
suppressed as $\sim{\rm e}^{-2n\theta_0}\;(n=1,2)$ yields a quadratic equation for $t=\tan(\omega_gU)$, 
\begin{equation}
a t^2 - 2b t + c > 0  \;\Longrightarrow \; 
 a(t-t_+)(t-t_-)>0,\quad 
t_\pm = \frac{1}{a}\big[b\pm\sqrt{\Delta}\big]\,,\qquad 
\Delta = b^2 - ac
,\;
\label{theta0 existence quadratic equation}
\end{equation}
where 
\begin{equation}
a= \frac{h_+^2\!+\!h_\times^2}{2}\left[-\Big(\!\theta_0^2\!-\!\theta_0+\frac12\Big)c_\psi\!+\!\frac12\right]
,\quad
b= \frac{h_+^2\!-\!h_\times^2}{2}\Big(\!\theta_0^2\!-\!\theta_0\!+\!\frac12\Big)s_\psi
,\quad
c= \frac{h_+^2\!+\!h_\times^2}{2}\left[\Big(\!\theta_0^2\!-\!\theta_0+\frac12\Big)c_\psi\!-\!\frac12\right]
,
\label{abc}
\end{equation}
and we made use of, $\cos(2\omega_gU)=(1-t^2)/(1+t^2)$ and  $\sin(2\omega_gU)=2t/(1+t^2)$.
The discriminant in Eq.~(\ref{theta0 existence quadratic equation})  is then,
\begin{eqnarray}
\Delta  \!\!&=&\!\!  \left(\frac{h_+^2\!+\!h_\times^2}{2}\right)^2
     \left[\Big(\!\theta_0^2\!-\!\theta_0+\frac12\Big)^2\!-\!\frac14\right]
\!-\!h_+^2h_\times^2 \Big(\!\theta_0^2\!-\!\theta_0\!+\!\frac12\Big)^2s^2_\psi
\nonumber\\
 \!\!&=&\!\!  \left(\frac{h_+^2\!-\!h_\times^2}{2}\right)^2
     \left[\Big(\!\theta_0^2\!-\!\theta_0+\frac12\Big)^2\!-\!\frac14\right]
\!+\!h_+^2h_\times^2 \left[\Big(\!\theta_0^2\!-\!\theta_0\!+\!\frac12\Big)^2c^2_\psi\!-\!\frac14\right]
\,.\qquad\;
\label{discriminant}
\end{eqnarray}
The roots $t_\pm$ in Eq.~(\ref{theta0 existence quadratic equation})  of the equation,
$a t^2 - 2b t + c = 0$, are real if $\Delta\geq 0$, from which we conclude that the inequality 
in Eq.~(\ref{theta0 existence quadratic equation}) is satisfied: 
\begin{enumerate}
\item[(1)] when $a>0 $ and $\Delta<0$: the inequality in Eq.~(\ref{theta0 existence quadratic equation}) 
is satisfied for all $U\in {\mathbb R}$;
\item[(2)]  when $a>0 $ and $\Delta\geq 0$: the inequality in Eq.~(\ref{theta0 existence quadratic equation}) 
is satisfied for $t=\tan(\omega_g U)<t_-$ and $\tan(\omega_g U)>t_+$;
\item[(3)]  when $a<0 $ and $\Delta\geq 0$: the inequality in Eq.~(\ref{theta0 existence quadratic equation})  
is satisfied for $t_-<\tan(\omega_g U)<t_+$;
\item[(4)]  when $a<0 $ and $\Delta<0$: the inequality in Eq.~(\ref{theta0 existence quadratic equation}) 
is never satisfied.
\end{enumerate}
From Eqs.~(\ref{abc}) and~(\ref{discriminant}) we then see that $a\lessgtr 0$ and $\Delta 
{\geq\atop<}$ imply, 
\begin{eqnarray}
\cos(\psi)  \!\!&\lessgtr&\!\! \frac{1}{2\theta_0(\theta_0-1)+1}\simeq \frac{1}{2\theta_0^2}
\label{conditions for cos psi}\\
 \cos^2(\psi)\!\!&{\geq\atop<}&\!\!  \frac{1}{[2\theta_0(\theta_0-1)+1]^2}
  - \frac14\left(\frac{h_+}{h_\times}\!-\!\frac{h_\times}{h_+}\right)^2
    \left[1-\frac{1}{[2\theta_0(\theta_0-1)+1]^2}\right]
 \simeq \frac{1}{4\theta_0^4} - \frac14\left(\frac{h_+}{h_\times}\!-\!\frac{h_\times}{h_+}\right)^{\!2}
, \qquad
 \nonumber
\end{eqnarray}
where the last inequalities represent good approximations when $\theta_0\gg 1$. Notice that if, 
\begin{equation}
 \left|\frac{h_+}{h_\times}\!-\!\frac{h_\times}{h_+}\right|>
       \frac{1}{\sqrt{\theta_0(\theta_0-1)(\theta_0^2-\theta_0+1)}}
      \simeq \frac{1}{\theta_0(\theta_0-1)}
\;\Longrightarrow \Delta > 0 
\,,
\label{condition for Delta > 0}
\end{equation}
and options (1) and (4) are absent, and an Unruh-DeWitt detector gets excited only during 
parts of the period. For circularly polarized gravitational waves, $\cos(\psi)=0$, and excitations occur 
when,
\begin{equation}
 t_-<\tan(\omega_g U)<t_+
 \,.
 \label{condition for cuts to occur}
\end{equation}

There is another important difference between detector's transition rate induced by 
circularly polarized and elliptically polarized gravitational waves. Upon rewriting Eq.~(\ref{function H2})
at leading order in ${\rm e}^{2\theta_0}$,
\begin{eqnarray}
\frac{h_+^2\!+\!h_\times^2}{2}
\!+\!\bigg[\frac{h_+^2\!+\!h_\times^2}{2}c_\psi\cos(2\omega_gU)\!-\!\frac{h_+^2\!-\!h_\times^2}{2}s_\psi\sin(2\omega_gU)\bigg]
\Big[2\theta_0(\theta_0\!-\!1)\!+\!1\Big]
\!+\!{\cal O}\left({\rm e}^{-2\theta_0}\right)
        =4\theta_0^2{\rm e}^{-2\theta_0}
\,,\qquad
\label{function H2: approx}
\end{eqnarray}
we see that only the first term in Eq.~(\ref{function H2: approx})
 survives in the circularly polarized case, in which $h_+=h_\times$ and $\cos(\psi)=0$.
Very close to that point, the first term $(h_+^2\!+\!h_\times^2)/{2}$ dominates, and 
the beginning of the cut is still well approximated in terms of the 
Lambert function as ({\it cf.} Eq.~(\ref{Lambert function approximation})), 
\begin{equation}
\theta_0(h_+,h_\times) \simeq 
- \Re\bigg[W\bigg(\!\!-\frac12\sqrt{\frac{h_+^2+h_\times^2}{2}}\,\bigg)
                                           \bigg]
\,,
\label{root: Lambert function}
\end{equation}
whose small $\bar h\equiv \sqrt{(h_+^2+h_\times^2)/2}-$expansion 
is highly nonanalytic. On the other hand, when
deviations from the circularly polarized case are significant, the solution changes to, 
\begin{equation}
\theta_0(h_+,h_\times) \simeq - \frac{1}{2}\ln
\bigg[\frac{h_+^2\!+\!h_\times^2}{4}c_\psi\cos(2\omega_gU)\!-\!\frac{h_+^2\!-\!h_\times^2}{4}s_\psi\sin(2\omega_gU)\bigg]
\,,
\label{root: Lambert function}
\end{equation}
which exists when the argument of the logarithm is positive, {\it i.e.} when 
$\tan(2\omega_gU)\lessgtr \Big(\frac{h_+^2+h_\times^2}{h_+^2-h_\times^2}\Big)\cot(\psi)$
for $(h_+^2-h_\times^2)\sin(\psi)\lessgtr 0$.

With these considerations in mind, one can obtain an approximate expression 
for the detector rate in the massless limit 
({\it cf.} Eq.~(\ref{transition rate: UdW detector: nonpol: analytic 2})),
\begin{eqnarray}
{\cal R}(U,\Delta E)    \!\!&\approx&\!\!
\frac{\hbar\sqrt{\tilde\gamma}P_v\omega_g}{2\pi^2}
                 \sqrt{\frac{P_v\omega_g}{\Delta E}}
      \left(\frac{h_+^2c_+^2(U)\!+\!h_\times^2c_\times^2(U)}{2}\!-\!\frac{h_+^2\!+\!h_\times^2}{4}
               \right)^\frac{\Delta E}{2P_v\omega_g}\Theta_+
\,,\qquad 
\label{transition rate: UdW detector: pol: analytic massless}
\end{eqnarray}
where $\Theta_+=1$ when the argument inside the brackets is positive
(which is identical the positivity requirement on the argument of the logarithm
in Eq.~(\ref{root: Lambert function})),
and $\Theta_+=0$ when it is negative.
A second perturbative approximation can be obtained by expanding the integrand 
in Eq.~(\ref{transition rate: UdW detector 2}) in powers of the gravitational strain.
Making use of Eq.~(\ref{deformation matrix: momentum space: pol 5}) 
and keeping, for simplicity, the quadratic order terms only, 
one obtains for the detector's transition rate in the massless scalar case,
\begin{eqnarray}
{\cal R}(\Delta E)    \!\!&=&\!\! \frac{\hbar\sqrt{\tilde \gamma}P_v\omega_g}{2\pi}
                    \Bigg\{\frac{h^{2}(U)}{6} \!\left(\!1\!-\!\frac{\Delta E}{2P_v\omega_g}\right)^3\!
\!+\!\frac{2h^{2}(U)\!-\!(h_+^2\!+\!h_\times^2)}{4} 
           \!\left(\!1\!-\!\frac{\Delta E}{2P_v\omega_g}\right)
    \frac{\Delta E}{2P_v\omega_g}
          \Bigg\}\Theta(2P_v\omega_g\!-\!\Delta E)
\nonumber\\
\!\!&+&\!\!{\cal O}(h_{+,\times}^4)
\label{transition rate: UdW detector: pol n=1}
\,,\qquad 
\end{eqnarray}
where $h^2(U)=h_+^2c_+^2(U)+h_\times^2c_\times^2(U)$, and 
we have dropped the quartic terms as they are significantly more complicated
than in the circularly polarized case in Eq.~(\ref{transition rate: UdW detector: nonpol n=1 and 2}).

\begin{figure}[h!]
\vskip -.2cm
\centerline{\hspace{.in}
\epsfig{file=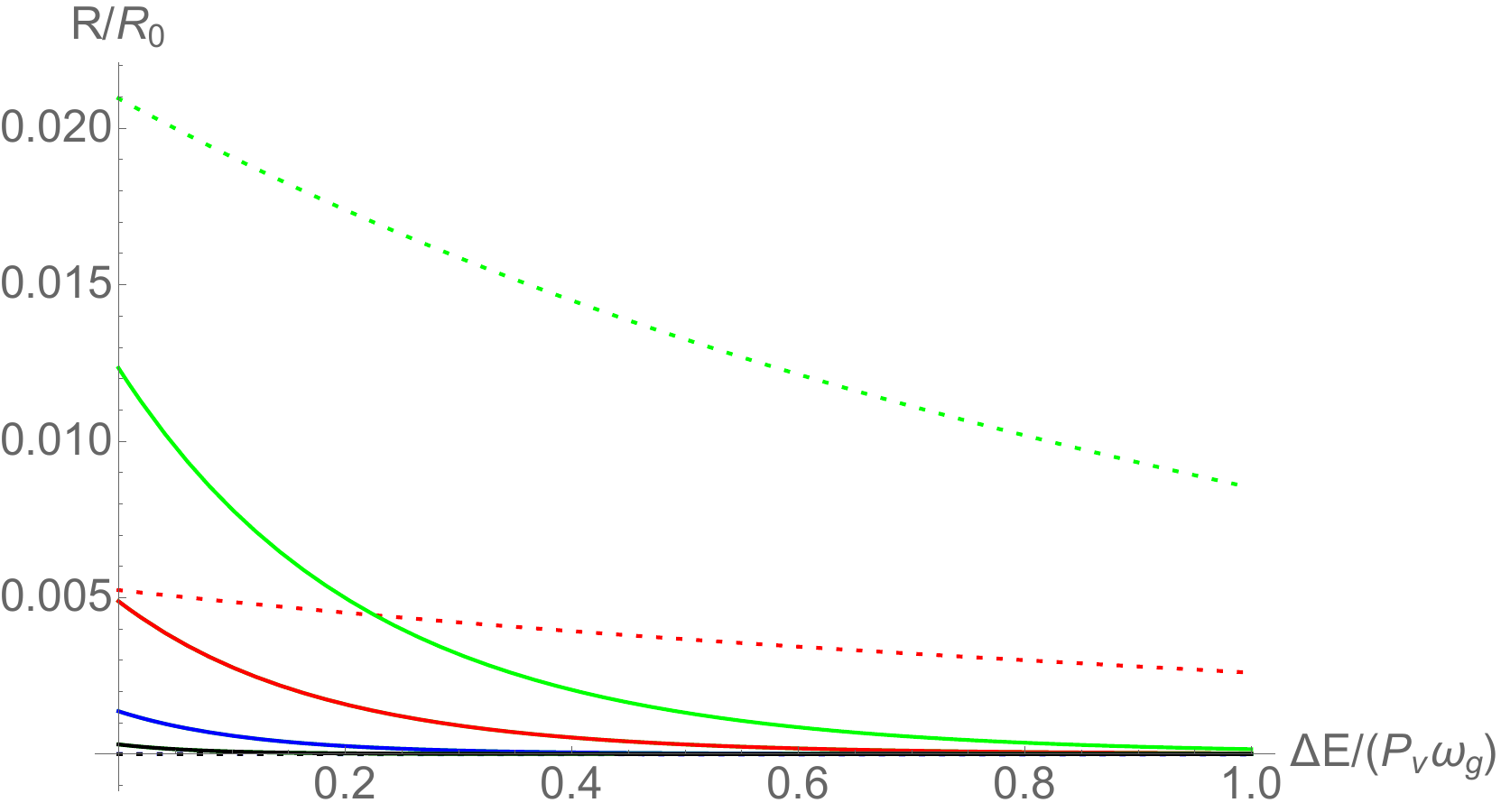, width=3.5in}
\hskip -0.2cm
\epsfig{file=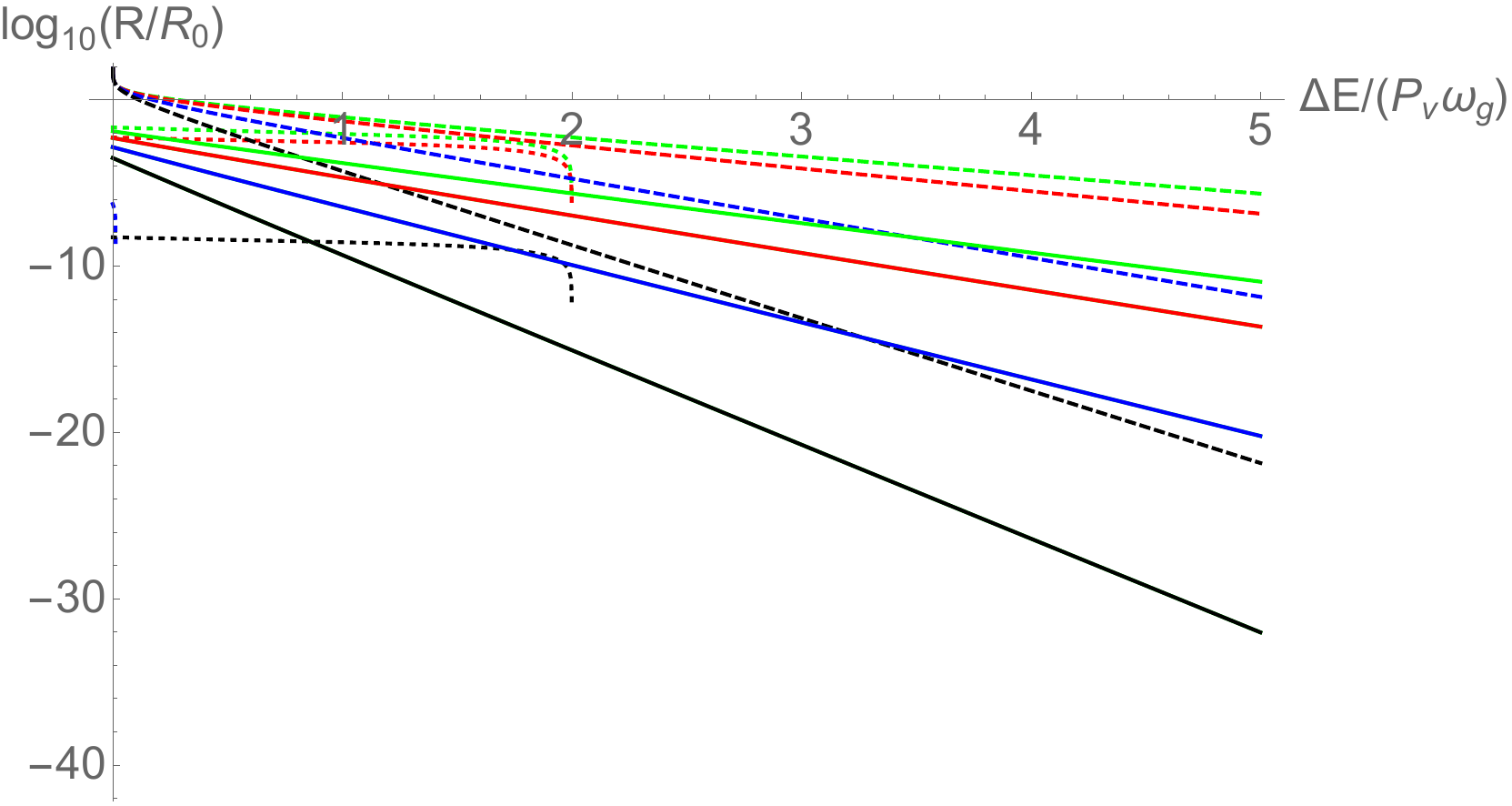, width=3.8in}
}
\vskip -0.1cm
\caption{\small {\it Left panel:} The detector's transition rate ${\cal R}$ for a massless scalar 
as a function of $\Delta E/(P_v\omega_g)$ 
in units of ${\cal R}_0=\hbar \sqrt{\tilde\gamma} P_v\omega_g/(2\pi ^2)$ 
as measured by a freely falling Unruh-DeWitt detector. All curves are for the moment in time, 
$U=\pi/4\omega_g$ and for the relative phase, $\psi=\pi/2$, which is the same phase difference 
as for circularly polarized waves. Different curves are for different values of 
the gravitational strains: $h_+=0.1$ and $h_\times=0.2$ (green),  
$h_+=0.01$ and $h_\times=0.1$ (red), $h_+=0.01$ and $h_\times=0.001$ (blue), 
and $h_+=0$ and $h_\times=0.0001$ (black).
{\it Right panel:} The same but now for $\log_{10}[{\cal R}/{\cal R}_0]$. 
The solid curves are numerical results, the dashed curves are obtained by using 
the approximation in Eq.~(\ref{transition rate: UdW detector: pol: analytic massless}),
and the dotted lines correspond to the approximation 
in Eq.~(\ref{transition rate: UdW detector: pol n=1}).
}
\label{figure six}
\end{figure}

In figure~\ref{figure six} we show selected numerical results of integrating
Eq.~(\ref{transition rate: UdW detector: pol B2}) in the massless limit, in which the transition rate reduces to
({\it cf.} Eq.~(\ref{transition rate: UdW detector: nonpol B3})),
\begin{eqnarray}
{\cal R}(\Delta E)    \!\!&=&\!\! \frac{\hbar\sqrt{\tilde\gamma}P_v\omega_g}
               {2\pi^2}\int_{\theta_0}^{\infty}
               \frac{{\rm d}\theta}{\theta\big[H^2(U,\theta)-\theta^2\big]^{1/2}} 
{\rm e}^{-\frac{\Delta E}{P_v\omega_g}\theta}
\,.\qquad 
\label{transition rate: UdW detector: pol B3}
\end{eqnarray}
In the same figure, for comparison, we also show the analytical estimates 
from Eq.~(\ref{transition rate: UdW detector: pol: analytic massless}) (dashed)
and from Eq.~(\ref{transition rate: UdW detector: pol n=1}) (dotted).
Notice that the latter approximation (obtained by expanding in powers of $h$) 
does not work as well as in the circularly polarized case, 
as it does not need to give a positive result in the whole interval, $0\leq \Delta E\leq 2P_v\omega_g$.~\footnote{The perturbative rate can be negative if $h^2(U)<(h_+^2+h_\times^2)/2$ and
when $x_-<\Delta E/(2P_v\omega_g)<{\rm min}[x_+,1]$, where $x_\pm$ are the roots of the 
equation, $x^2+2\big[2-3(h_+^2+h_\times^2)/(2h^2(U))\big]x+1=0$.}
The detector's transition rate for a massive scalar field can be studied analogously,
and therefore we leave it as an exercise.

\bigskip

None of the results presented in this section 
can be compared 
with those in Ref.~\cite{Chen:2021bcw}, where approximations 
were used which do not capture the effects of the cuts in figure~\ref{figure one}.~\footnote{Here 
we do not take account of 
the detector's rate induced by transitions to lower energy states
considered in Ref.~\cite{Chen:2021bcw}, for which $\Delta E<0$, as those 
are absent when the detector is in its ground state, and moreover such transitions would be hard 
to resolve from detector's response in Minkowski vacuum. The transitions we consider excite the 
detector, $\Delta E>0$, and they are completely absent in Minkowski vacuum.}

 \bigskip
 
 \section{Conclusions and outlook}
\label{Conclusions and outlook}

In section~\ref{Scalar propagator} we generalize the massive real scalar field propagator of 
Ref.~\cite{vanHaasteren:2022agf} to general gravitational waves propagating in one direction,
thus relaxing the monochromatic approximation used in Ref.~\cite{vanHaasteren:2022agf}.
The Wightman two-point functions are given in Eq.~(\ref{Wightman functions: general Lor viol solution})
and the propagator in Eqs.~(\ref{Feynman propagator: lightcone coordinates}--\ref{Feynman propagator: Cartesian coordinates}). We then show that the generalized propagator 
produces the one-loop results which are identical in form to the ones
obtained in Ref.~\cite{vanHaasteren:2022agf}.

In section~\ref{Unruh-DeWitt detector} we then study how a freely falling Unruh-DeWitt detector 
(which couples to a massless or massive scalar field) responds to 
the gravitational wave background.
We find that the deformation of the invariant distance induced by the gravitational waves 
gets fully compensated by the motion of a freely falling detector,
\footnote{
From Eq.~(\ref{Lorentz breaking distance: general}) one sees that for timelike distances 
$\Delta \bar x^2(x;x')
\rightarrow -(\Delta \tau\!-\!i\epsilon)^2$, such that in the classical limit (when 
$\epsilon\rightarrow 0$) $\Delta \bar x^2(x;x')$ reduces to the geodesic distance 
between points $x$ and $x'$, also known as the worldline. When the distance is lightlike or 
spacelike however, there is no classical analogue for $\Delta \bar x^2(x;x')$.
}
 thus leaving the effect of
the modified amplitude of vacuum fluctuations expressed by the $(u,u')-$dependent prefactor in 
Eq.~(\ref{Wightman functions: general Lor viol solution}), the effects of which we study in some detail. 
In this work we focus on studying detector's excitation rate induced by passing
gravitational waves, {\it i.e.} the rate of transitions from the detector's 
ground state (with energy $E_0$)
to an excited state $E$, for which $\Delta E=E-E_0>0$. These transitions are of a
particular interest as they are completely absent in Minkowski space, and therefore 
-- no matter how small it may be -- any observed rate 
can be interpreted as the detection of gravitational waves.

\bigskip 
We treat the detector's transition rate in three different approximations:
\begin{enumerate} 

\item[$\bullet$] {\it Numerical solution}, which can be considered to be exact. 
The resulting detector's excitation rate ${\cal R}$
is exponentially suppressed, and can be approximated by,
 ${\cal R}= {\cal R}_0(\Delta E,m,\omega_g)\exp\left[-\frac{\Delta E}{P_v\omega_g}\theta_0(h)\right]$,
 where $\theta_0(h)>0$ is the beginning of the cut in the complex $\Delta u$ plane, 
 whose functional dependence on $h$ is
 determined by Eq.~(\ref{cuts of Delta tau: nonpol}), and it is highly 
 nonanalytic;
$ {\cal R}_0(\Delta E,m,\omega_g)$ is a weak function of $\Delta E/(P_v\omega_g)$ and exponentially decays
 with increasing $m/(P_v\omega_g)$.

\item[$\bullet$] Expanding in powers of $h^2$, where $h$ is the gravitational wave strain. This generates
 a series of poles of the order $2n+2$, each of which contributes to the term $\sim h^{2n}$ 
when $0\leq\Delta E\leq  2n P_v\omega_g$, the first two contributions are shown in 
Eq.~(\ref{transition rate: UdW detector: nonpol n=1 and 2}) 
for circularly polarized gravitational waves and in Eq.~(\ref{transition rate: UdW detector: pol n=1}) for general elliptically polarized gravitational waves. 

The field theoretic interpretation of these contributions
is that they actively contribute when the scalar field absorbs $2n$ gravitons, each of which with energy
$P_v\omega_g$, such that the detector's energy can increase by $\Delta E\leq 2n P_v\omega_g$.
Figures~\ref{figure three} and~\ref{figure six} 
show that expanding in powers of the gravitational wave strain captures correctly 
the qualitative trend of the numerical solution, but at the quantitative level this 
approximation performs quite poorly. 
\item[$\bullet$] Expanding around the cut at $\theta=\theta_0(h)$ in Eqs.~(\ref{cuts of Delta tau: nonpol}) and~(\ref{function H2})
for circularly polarized and elliptically polarized gravitational waves, respectively.
The leading order result is shown in Eq.~(\ref{transition rate: UdW detector: nonpol: analytic 2})
for circularly polarized waves 
and in Eq.~(\ref{transition rate: UdW detector: pol: analytic massless}) for elliptically polarized waves. 
This approximation captures correctly the exponential decay  
of the detector's excitation rate with increasing $\Delta E>0$, 
that is its nonanalytic structure in the gravitational wave strain, but it fails to correctly model 
the exponential prefactor, as can be clearly seen from 
figures~\ref{figure three}, \ref{figure four}, \ref{figure five} and~\ref{figure six}.
\end{enumerate} 

In this work we have addressed the response of an Unruh-DeWitt detector to monochromatic, 
unidirectional, circularly polarized and elliptically polarized, gravitational waves. A more general investigation 
is warranted by relaxing any of the above mentioned restrictions. It would be, in particular, of interest to 
calculate the transition rate of the detector induced by a stochastic gravitational wave background. But to do that 
properly requires knowledge of the corresponding propagator, which is presently unknown.

\section{Appendices}
\label{Appendices}

\section*{Appendix A: Derivation of the Wightman functions}
\label{Appendix A: The method of mode sums} 

Here we briefly present a derivation of the Wightman functions by 
the method of mode sums.
Upon expanding the scalar field operator in terms of the momentum space 
mode functions $\phi_\pm(u,\vec k)$ 
and the creation and annihilation operators $\hat a^\dagger(\vec k)$ 
and $\hat a(\vec k)$,
 \begin{eqnarray}
  \hat\phi(u,\vec x_\perp,v) 
  \!\!&=&\!\! \int \frac{{\rm d}^{D-2}k_\perp {\rm d}k^{D-1}}{(2\pi)^{D-1}}
     {\rm e}^{i\vec k_\perp\cdot\vec x_\perp}
      \Big[{\rm e}^{-\frac{i}{2}\Omega_-(\vec k\,) v}\phi_+(u,\vec k\,)\hat a(\vec k\,)
       + {\rm e}^{\frac{i}{2}\Omega_+(\vec k\,) v}\phi_-(u,\vec k\,)\hat a^\dagger(-\vec k\,)\Big]
\,,
\label{field expansion}
\end{eqnarray}
which obey a standard non-vanishing commutation relation,
\begin{equation}
\left[\hat a(\vec k\,),\hat a^\dagger(\vec k')\right]
=(2\pi)^{D-1}\delta^{D-1}(\vec k\!-\!\vec k')
\,,
\nonumber
\end{equation}
one obtains the following expressions for the positive and negative frequency Wightman functions,
\begin{eqnarray}
i\Delta^{(+)}(x;x') \!\!&=&\!\! \left\langle\Omega\right|\hat\phi(x)\hat\phi(x')\left|\Omega\right\rangle = 
 \int\! \frac{{\rm d}^{D-1}k}{(2\pi)^{D-1}}\,
{\rm e}^{i \vec k_\perp\cdot\,\Delta \vec x_\perp 
 -\frac{i}{2}\Omega_-(\vec k)\Delta v}
\phi_+(u,\vec k\,)\phi_-(u',-\vec k\,)
\,,
 \label{positive frequency Wightman function}\\
 i\Delta^{(-)}(x;x') \!\!&=&\!\! \left\langle\Omega\right|\hat\phi(x')\hat\phi(x)\left|\Omega\right\rangle 
                          =
\int\!  \frac{{\rm d}^{D-1}k}{(2\pi)^{D-1}}\,
{\rm e}^{i \vec k_\perp\cdot\,\Delta \vec x_\perp 
 +\frac{i}{2}\Omega_+(\vec k)\Delta v}
\phi_-(u,\vec k\,)\phi_+(u',-\vec k\,)
\nonumber\\
\!\!&& \hskip 3cm
=\, \int\!  \frac{{\rm d}^{D-1}k}{(2\pi)^{D-1}}\,
{\rm e}^{-i \vec k_\perp\cdot \,\Delta \vec x_\perp 
             +\frac{i}{2}\Omega_-(\vec k)\Delta v}
\phi_-(u,-\vec k\,)\phi_+(u',\vec k\,)
\,,\qquad\;\;
\label{negative frequency Wightman function}
\end{eqnarray}
where $\Delta \vec x_\perp = \vec x_\perp-\vec x_\perp^{\,\prime}$, $\Delta v = v\!-\!v'$,
and we have assumed,
\begin{equation}
\hat a(\vec k\,)|\Omega\rangle = 0
\,,\qquad (\forall \vec k\in {\mathbb R}^{D-1})
\,.
\label{def: vacuum state}
\end{equation}
%
The functions $\phi_+(u,\vec k\,)$ and $\phi_-(u',\vec k\,)$ are the 
positive and negative frequency mode functions obeying,
\begin{equation}
   \!\left(\!\partial_u
     \!\pm\! \frac{i}{2\Omega_\mp}\Big[\big(g^{ij}(u)-\delta^{ij}_\perp\big)k_ik_j\Big]
      \!\pm\! \frac{i}{2}\Omega_\pm\right)\!
       \big[\gamma^{1/4}(u)\phi_\pm(u,\vec k)\big] = 0
,\hskip 0.2cm
\label{EOM scalar in 4 mode functions}
\end{equation}
where $\Omega_\pm(\vec k) = \omega\pm k^{D-1}$, 
$\gamma(u)={\rm det}[g_{ij}(u)]$, 
and $\omega=\sqrt{\|\vec k\|^2+m^2}$.
This is a first order differential equation 
in $u$ and therefore can be easily solved. The properly normalized, ground
state solutions of 
Eq.~(\ref{EOM scalar in 4 mode functions}) are given by, 
\begin{equation}
\phi_\pm(u,\vec k) = \frac{1}{\gamma^{1/4}(u)}
 \sqrt{\frac{\hbar}{2\omega}}
\exp\left[\mp\frac{i\Omega_\pm }{2}\left(u 
  +\frac{k_i k_j}{\omega^2_\perp}\int^u {\rm d}\bar u \big(g^{ij}(\bar u) - \delta^{ij}\big)\right)\right]
\,,\qquad
\label{appendix A: mode functions}
\end{equation} 
where $\omega_\perp^2 = \omega^2 - (k^{D-1})^2$
and $\omega^2 = \|\vec k\|^2+m^2$. The factor $\gamma^{-1/4}(u)$
in the normalization of the mode functions~(\ref{appendix A: mode functions})
can be traced back to the Wronskian condition for the mode functions (which in turn  
originates from canonical quantization).
Together with the condition~(\ref{def: vacuum state}), the choice of pure positive 
(negative) frequency solutions for the functions $\phi_+$ ($\phi_-$) 
in Eq.~(\ref{field expansion}) uniquely specify the Gaussian state of the system,
which we consider the vacuum state.
More general (pure) Gaussian states can be obtained by exacting a Bogolyubov 
transformation on the operators $\hat a(\vec k)$ and $\hat a^\dagger(\vec k)$.
However, these states are excited states of the system, in the sense that their energy 
per mode is higher 
than the energy of the state $|\Omega\rangle$ defined 
in Eq.~(\ref{def: vacuum state}). In that respect the state $|\Omega\rangle$ 
used here to quantize the scalar field and calculate the Wightman functions
can be considered as the vacuum state.

Upon inserting the mode functions~(\ref{appendix A: mode functions}) 
into Eq.~(\ref{positive frequency Wightman function})
and converting ${\rm d}k^{D-1}$ into ${\rm d}\Omega_+$ 
one obtains, 
\begin{eqnarray}
i\Delta^{(+)}(x;x') \!\!&=&\!\!\frac{\hbar}{[\gamma(u)\gamma(u')]^{1/4}} 
          \int \!\frac{{\rm d}^{D-2} k_\perp}{2(2\pi)^{D-1}}
    {\rm e}^{i\vec k_\perp\cdot  \Delta\vec x_\perp}
     \! \int_0^\infty \! \frac{{\rm d} \Omega_+}{\Omega_+}
\label{propagator integral D: 1}\\
\!\!&&\!\!\!\!\!\!\!\!\!\!\!\!\!\!\times
 \exp\bigg\{\!-\frac{i}{2} 
 \bigg[\Omega_+\bigg(\!1\!+\!\frac{({\cal G}^{xx}(u;u')\!-\!1) k_x^2
               \!+\!({\cal G}^{yy}(u;u')\!-\!1) k_y^2
                         \!+\!2{\cal G}^{xy}(u;u') k_xk_y}{\omega_\perp^2}\bigg)
                \Delta u 
                              \!+\! \frac{\omega_\perp^2}{\Omega_+}\Delta v\bigg]
         \bigg\}
,\quad
\nonumber
\end{eqnarray}
where 
\begin{equation}
{\cal G}^{ij}(u;u') = \frac{1}{\Delta u}\int_{u'}^ug^{ij}(\bar u)\, {\rm d}\bar u
                                \,,\qquad (\Delta u =u\!-\!u')
\,.
\label{G ij matrix}
\end{equation}
One can evaluate the $\Omega_+$-integral in Eq.~(\ref{propagator integral D: 1})
by making use of Eq.~(3.471.9) in Ref.~\cite{Gradshteyn:2014},
\begin{equation}
i\Delta^{(+)}(x;x') = \frac{\hbar}{[\gamma(u)\gamma(u')]^{1/4}}
\int\!\frac{{\rm d}^{D-2} k_\perp}{(2\pi)^{D-1}}
   {\rm e}^{i\vec k_\perp\cdot  \Delta\vec x_\perp}\!
                     K_0\Bigg(\omega_\perp\sqrt{-\Delta u_\epsilon\Delta v_\epsilon
                     \bigg(\!1\!+\!\frac{{\cal G}^{xx} k_x^2\!+\!{\cal G}^{yy}k_y^2
                         \!+\!2{\cal G}^{xy} k_xk_y}{\omega_\perp^2}\bigg)}\,\Bigg)
,\quad
\label{propagator integral D: 2}
\end{equation}
where we have introduced a shorthand notation for 
the $i\epsilon$ prescriptions, $\Delta u_\epsilon\equiv \Delta u-i\epsilon$ 
and $\Delta v_\epsilon\equiv \Delta v-i\epsilon$. The argument squared of the 
modified Bessel function of the second kind $K_0$ is a quadratic form
$Q(k_x,k_y)$ which can be diagonalized by a simple 
${\cal G}^{ij}$-dependent rotation. For that purpose 
the eigenvalues and determinant of ${\cal G}^{ij}$ are useful, 
 \begin{eqnarray}
 {\cal G}_\pm \!\! &=&\!\! \frac12\left({\cal G}^{xx} +{\cal G}^{yy}\right) \pm
    \sqrt{\frac14\left({\cal G}^{xx}-{\cal G}^{yy}\right) - \big({\cal G}^{xy}\big)^2}
\label{Appendix A: eingenvalues Q}\\
{\cal G}^{-1}(u;u')\!\! &\equiv&\!\!{\rm det}\big[{\cal G}^{ij}\big] = {\cal G}_+{\cal G}_-
  ={\cal G}^{xx}{\cal G}^{yy}- \big({\cal G}^{xy}\big)^2 
  \,.
\label{Appendix A: determinant G ij}
\end{eqnarray}
Upon rotating the momenta $(k_x,k_y)$ into the diagonal frame $(\tilde k_x,\tilde k_y)$ and 
renormalizing them as, 
$(\bar k_x,\bar k_y)=(\sqrt{{\cal G}_+}\tilde k_x,\sqrt{{\cal G}_-}\tilde k_y)$, 
Eq.~(\ref{propagator integral D: 2}) reduces to,
\begin{eqnarray}
i\Delta^{(+)}(x;x') =\frac{1}{[\gamma(u)\gamma(u')]^{1/4}\sqrt{{\rm det}[{\cal G}^{ij}]}} \int\!\frac{{\rm d}^{D-2} {\bar k}_\perp}{(2\pi)^{D-1}}
   {\rm e}^{i\vec {\bar k}_\perp\cdot  \Delta\vec {\bar x}_\perp}\!
                     K_0\Big(\bar\omega_\perp\sqrt{-\Delta u_\epsilon\Delta v_\epsilon}\,\Big)
,\quad
\label{propagator integral D: 3}
\end{eqnarray}
where $\bar\omega_\perp^2=\bar k_x^2+\bar k_y^2+\sum_{n=3}^{D-2}k_n^2+m^2$
(the sum contributes when $D>4$)
and $\Delta \vec {\bar x}$ is defined by,
 $\vec {k}_\perp\cdot \Delta \vec {x}=\vec {\bar k}_\perp\cdot \Delta \vec {\bar x}$.
By expressing ${\rm d}^{D-2}k$ in spherical coordinates and integrating over the angles, 
Eq.~(\ref{propagator integral D: 3}) can be recast as, 
\begin{eqnarray}
i\Delta^{(+)}(x;x') 
 =\frac{\hbar}{[\gamma(u)\gamma(u')]^{1/4}\sqrt{{\rm det}[{\cal G}^{ij}]}}
\frac{1}{ \|\Delta\vec{\bar x}_\perp\|^\frac{D-4}{2}}
\int_0^\infty\!\! \frac{{\rm d}\bar k_\perp}{(2\pi)^\frac{D}{2}} \bar k_\perp^\frac{D-2}{2}
              J_\frac{D-4}{2}\big(\bar k_\perp \|\Delta\vec{\bar x}_\perp\|\big)
               K_0\big(\bar\omega_\perp\sqrt{\!-\!\Delta u_\epsilon\Delta v_\epsilon}\,\big)
\,,
\nonumber\\
\label{propagator integral D: 4}
\end{eqnarray}
The final integral over $\bar k_\perp$  can be evaluated by using (6.596.7)
 in Ref.~\cite{Gradshteyn:2014},
\begin{eqnarray}
i\Delta^{(+)}(x;x')  \!\!\!&=&\!\! 
\frac{\hbar m^{D-2}}{(2\pi)^{D/2}[\gamma(u)\gamma(u')]^{1/4}
    \sqrt{{\rm det}[{\cal G}^{ij}](u;u')}}
\frac{ K_{\frac{D-2}{2}}\bigg(m\sqrt{\!-(\Delta u\!-\!i\epsilon)(\Delta \!-\!i\epsilon)
                      \!+\!\|\Delta\vec{\bar x}_\perp\|^2}\,\bigg)}
                      {\left(m\sqrt{\!-(\Delta u\!-\!i\epsilon)(\Delta v\!-\!i\epsilon)\!
                         +\!\|\Delta \vec{\bar x}_\perp\|^2}\,\right)^\frac{D-2}{2}}
\,,\qquad\;\;
\label{propagator integral D: 5}
\end{eqnarray}
where we have restored the original $i\epsilon$ prescriptions and where,
\begin{equation}
\|\Delta \vec{\bar x}_\perp\|^2 =\sum_{i,j=1,2}\Delta x^i{\cal G}_{ij}(u; u')\Delta x^j
   + \sum_{n=3}^{D-2}\Delta x_n^2
\,,
\label{propagator integral D: 6}
\end{equation}
where  the sum contributes when $D>4$.
${\cal G}_{ij}(u; u')$ is the matrix inverse of ${\cal G}^{ij}(u; u')$,
${\cal G}^{ik}(u; u'){\cal G}_{kj}(u; u')=\delta^{i}_{\;j}$, such that:
\begin{equation}
 {\cal G}(u;u') \equiv {\rm det}[{\cal G}_{ij}(u;u')]=\frac{1}{{\rm det}[{\cal G}^{ij}]}
 \,.
 \end{equation}
The negative frequency Wightman function~(\ref{negative frequency Wightman function})
satisfies, 
$ i\Delta^{(-)}(x';x) = i\Delta^{(+)}(x;x')$ and 
\begin{equation}
i\Delta^{(-)}(x;x') = \left[ i\Delta^{(+)}(x;x')\right]^*
\,, 
\label{propagator integral D: 7}
\end{equation}
and therefore $ i\Delta^{(-)}(x;x')$ can 
be obtained by taking a complex conjugate of Eq.~(\ref{propagator integral D: 5}),
reproducing the Wightman functions in 
Eq.~(\ref{Wightman functions: general Lor viol solution}).

\section*{Appendix B: Inverse quartic root cuts}
\label{Appendix B: Inverse quartic root cuts} 

In the case of polarized gravitational waves, the cut structure in the complex plane may be 
richer than the one shown in figure~\ref{figure one}. To see that, let us analyse
the term $1/[\gamma(u)\gamma(u')]^{1/4}$ in Eq.~(\ref{transition rate: UdW detector 2}) 
in the complex $\Delta u$-plane.
For simplicity we consider here maximally polarized gravitational waves. Let us begin with
the $+$ polarization. Recalling that $u=U+\Delta u/2$ and $u'=U-\Delta u/2$ one can write,
\begin{eqnarray}
\gamma(u) = 1-h_+^2\cos^2(\omega_g u)
                  = 1-\frac{h_+^2}{2}\big[1+\cos(2\omega_g U)\cos(\omega_g\Delta u)
                            -\sin(2\omega_g U)\sin(\omega_g\Delta u) \big]
\,,
\nonumber
\end{eqnarray}
from which we infer,
\begin{eqnarray}
\!\gamma(u)\gamma(u')\! = \!
 1\!-\!h_+^2\big[1\!+\!\cos(2\omega_g U)\cos(\omega_g\Delta u)
                            \big]
\!+\!\frac{h_+^4}{4} \big[\big(1\!+\!\cos(2\omega_g U)\cos(\omega_g\Delta u)\big)^2
                            \!\!-\sin^2(2\omega_g U)\sin^2(\omega_g\Delta u)
                            \big]
.\nonumber
\end{eqnarray}
To see that there are poles in the complex $\Delta u$ plane, let us 
transform this equation
into the variables,
$\omega_g\Delta u \rightarrow 2i\theta - 2\zeta$,
\begin{eqnarray}
\cosh^2(2\theta + 2i\zeta) - 2\left(\frac{2}{h_+^2}-1\right)\cos(2\omega_g U)\cosh(2\theta + 2i\zeta)
          +\left(\frac{4}{h_+^4}-\frac{4}{h_+^2}+\cos(2\omega_g U)\right)        = 0
\,.
\label{appendix B: pole equation}
\end{eqnarray}
The roots of this quadratic equation are given by,
\begin{eqnarray}
\left[\cosh(2\theta + 2i\zeta)\right]_\pm =  \left(\frac{2}{h_+^2}-1\right)\cos(2\omega_g U)
          \pm i\frac{2}{h_+}\sqrt{\frac{1}{h_+^2}-1}\sin(2\omega_g U) 
\,,
\label{appendix B: pole equation: roots}
\end{eqnarray}
or equivalently, 
\begin{eqnarray}
\cosh(2\theta) \!\!&=&\!\!  \frac{2}{h_+^2} - 1
\;\Longrightarrow \; 
\theta_\pm = \frac12\ln\left[\frac{2}{h_+^2}-1\pm\frac{2}{h_+}\sqrt{\frac{1}{h_+^2}-1}
                                            \right]
       = \pm\frac12\ln\left[\frac{2}{h_+^2}-1+\frac{2}{h_+}\sqrt{\frac{1}{h_+^2}-1}
                                            \right]
\,,\qquad 
\nonumber\\
(\zeta_\pm)_n  \!\!&=&\!\!  \pm\omega_g U + \pi n\,,\qquad (n\in \mathbb{Z})
\,.
\label{appendix B: pole equation: roots 2}
\end{eqnarray}
The poles in the case of $\times$-polarized gravitational waves are given by simply replacing 
$h_+\rightarrow h_\times$.  The result in Eq.~(\ref{appendix B: pole equation: roots 2}) 
shows that singly polarized gravitational waves generate a richer structure of cuts in the complex $\Delta u$ plane. 
In particular, the term $1/[\gamma(u)\gamma(u')]^{1/4}$ generates four of its own inverse 
quartic root cuts (for $n=0$), 
whose real parts `walk' along the real axis as, $\Re[\Delta u] = \pm 2U$, 
and the cuts begin at, $\Im[\Delta u] = \theta_\pm/\omega_g$. 
In this work we not attempt to model 
the detector's excitation rate due to these cuts.

\end{document}